\documentclass[aps,pra,twocolumn,notitlepage,showpacs,superscriptaddress,nofootinbib]{revtex4-1}
\usepackage{graphicx}
\usepackage{bm}
\usepackage{amsmath}
\usepackage[usenames,dvipsnames]{color}
\usepackage{amsfonts}
\usepackage[colorlinks]{hyperref}
\usepackage[english]{babel}
\usepackage[normalem]{ulem}

\date{\vspace{-5ex}}

\newcommand\redout{\bgroup\markoverwith
{\textcolor{red}{\rule[0.5ex]{2pt}{0.8pt}}}\ULon}

\newcommand{\ket}[1]{\left|{#1}\right\rangle}

\newcommand{\ccc}{\underline{\mathbf{\hat{c}}}}
\newcommand{\cc}{\mathbf{\hat{c}}}
\newcommand{\Htheta}{\text{\^H}_\theta}
\newcommand{\Hs}{\text{\^H}_{S}}
\newcommand{\CCC}{\text{\^C}_\theta}
\newcommand{\UU}{\text{\^U}}

\begin{document}

\title{Topological bound states of a quantum walk with cold atoms}
\author{Samuel Mugel}
\affiliation{Mathematical Sciences, University of Southampton, Highfield, Southampton, SO17 1BJ, United Kingdom}
\affiliation{ICFO-Institut de Ciencies Fotoniques, The Barcelona Institute of Science and Technology, 08860 Castelldefels (Barcelona), Spain}
\author{Alessio Celi}
\affiliation{ICFO-Institut de Ciencies Fotoniques, The Barcelona Institute of Science and Technology, 08860 Castelldefels (Barcelona), Spain}
\author{Pietro Massignan}
\affiliation{ICFO-Institut de Ciencies Fotoniques, The Barcelona Institute of Science and Technology, 08860 Castelldefels (Barcelona), Spain}
\author{J\'{a}nos K.\ Asb\'{o}th}
\affiliation{Institute for Solid State Physics and Optics, Wigner Research Centre,
Hungarian Academy of Sciences, H-1525 Budapest P.O. Box 49, Hungary}
\author{Maciej Lewenstein}
\affiliation{ICFO-Institut de Ciencies Fotoniques, The Barcelona Institute of Science and Technology, 08860 Castelldefels (Barcelona), Spain}
\affiliation{ICREA-Instituci\'{o} Catalana de Recerca i Estudis Avan\c{c}ats, E-08010 Barcelona, Spain}
\author{Carlos Lobo}
\affiliation{Mathematical Sciences, University of Southampton, Highfield, Southampton, SO17 1BJ, United Kingdom}
\date{\today}

\begin{abstract}
We suggest a method for engineering a quantum walk, with cold atoms as walkers, which presents topologically non-trivial properties. We derive the phase diagram, and show that we are able to produce a boundary between topologically distinct phases using the finite beam width of the applied lasers. A topologically protected bound state can then be observed, which is pinned to the interface and is robust to perturbations. We show that it is possible to identify this bound state by averaging over spin sensitive measures of the atom's position, based on the spin distribution that these states display. Interestingly, there exists a parameter regime in which our system maps on to the Creutz ladder.
\end{abstract}

\pacs{03.65.Vf, 37.10.Jk, 67.85.-d}

\maketitle

\section{Introduction}
\label{sec:Intro}

Random walks have found extensive use in modelling intrinsically random systems as well as for designing computer algorithms. The quantum analogue of the random walk, the quantum walk, is obtained by replacing the walker by a quantum particle, where path interference effects result in a myriad of new properties \cite{Chisaki2012} and make quantum walks relevant to quantum algorithms \cite{Childs2003, Ambainis04quantumwalk, Magniez2005, Farhi07aquantum}.

Beyond quantum algorithms, the study of quantum walks has been motivated by the study of fundamental phenomena. They were found to be intimately related to the path integral formalism \cite{Feynman2012} and the Dirac equation \cite{Strauch2006}. It was even found that some many body systems can be well modelled using a quantum walk on a one dimensional (1D) semi-infinite lattice, the sites of which represent the system's energy levels \cite{Oka2005}. Experimental implementations of quantum walks have been realised with photons \cite{Peruzzo2010, Schreiber2011, Schreiber2012, Kitagawa2012a, Crespi2013, Rechtsman2013, Cardano2015, Cardano2015a}, and single and multiple cold atoms \cite{Karski2009, Genske2013, Preiss2014, Robens2015, Robens2015a, Meier2016} and ions \cite{Schmitz2009, Zahringer2010}, which were relevant to the study of Anderson localisation, decoherence and reversibility in strongly interacting systems.

Despite their simplicity, quantum walks present rich topological phenomena \cite{Rudner2009, Kitagawa2010a, Rapedius2012}, as they can realise all known topological classes in one and two dimensions \cite{Schnyder2008}. For a detailed explanation of the emergence of topological phenomena, we refer the reader to Ref.\ \cite{asboth2016short}. Discrete time quantum walks, being periodically driven systems, can present topological invariants which are not found in the topological classification of Hamiltonians, as was shown in \cite{Kitagawa2010, Asboth2013}. Thanks to these properties, they form an ideal platform for realising Floquet bound states, which are bound as a result of the system's dynamics. These states further differ from bound states in static systems by presenting constrained dynamics when considered at half time-steps \cite{Asboth2014}. The topological properties of quantum walks and of Floquet bound states have been explored by means of photonic experiments in one and two dimensions \cite{Kitagawa2012a, Rechtsman2013, Cardano2015, Cardano2015a}.

In this paper, we suggest an experimental scheme for realising a quantum walk with cold atoms in a 1D optical lattice, and show that this system is topologically non-trivial. 
Here and in the following, we define a time-step in a discrete time quantum walk as the sequence of a translation operation, which transports a right- and left-walker in opposite directions, and of a coin operation, which brings the system into a superposition of right- and left-walkers.
Cold atomic gases appear as a natural candidate for this type of application, as it is possible to exert extremely fine control over them. Additionally, cold atoms suffer from few losses relative to photons, and optical lattices have scalable size, allowing for a very long evolution with many time-steps. It was shown in Refs.\ \cite{Ruostekoski2008, Jiang2011, Leder2016a} that 1D atomic gases are a powerful tool for generating topological bound states in static systems, or for building topologically protected edge states by using a synthetic dimension \cite{Celi2013, Mancini, Stuhl2015}. Inspired by these results, we suggest a method to generate Floquet bound states by spatially controlling the system's parameters. We show that these topologically protected states have a heavily constrained spin distribution. Thanks to this property, a spin sensitive measure of the system's probability distribution is sufficient to identify the system's topological bound states. This information can be retrieved by averaging over measurements of individual atom positions at a time $t$. This can be done in a single measurement, by performing the experiment multiple times in an array of 1D tubes, then observing the atoms' position using e.g: a quantum gas microscope. It would also become possible to study the Floquet bound states' robustness to interactions and to faults in the periodic driving, thereby providing useful information on systems which are still poorly understood.

In this paper's second section, we suggest an experimental protocol to realise a quantum walk with a single, two-state atom in a 1D optical lattice. The idea is to use the particle's internal degree of freedom to bring it into a superposition of going right and left simultaneously, which we do by driving the system periodically with a spin mixing operation. In Sec.\ \ref{sec:Model}, the equations governing the time evolution are presented. We use these operators to perform numerical simulations of the system, and show that the atoms have the dynamics of a quantum walk. In Sec.\  \ref{sec:TopoProp}, we show that this system is topologically non-trivial and derive its phase diagram. We find that the topological phase can be changed by changing the spin mixing angle, allowing us to generate a topological boundary. In Sec.\ \ref{sec:BoundState}, we populate the bound state that appears at this interface, and suggest a method for measuring its presence. In Sec.\ \ref{sec:PairOfJR}, we consider an interesting limit of this Hamiltonian, which also presents bound states, despite being topologically trivial. We explain this by showing that these are in fact a pair of Jackiw-Rebbi states, and study the mechanism according to which they can hybridise. In Appendix \ref{sec:MapToCreutz}, we show that, in a certain parameter regime, the system maps onto the Creutz ladder. In Appendix \ref{sec:DoubleQW}, we show that the protocol we suggest can be understood as a superposition of two independent quantum walks. In Appendix \ref{sec:Symmetries}, we explain the method used to find the system's symmetries.

\section{Experimental proposal}
\label{sec:Exp}

Our idea to implement a quantum walk is as follows: particles are allowed to evolve in a medium which accommodates right movers and left movers. By periodically applying a pulsed operation which has amplitude to interconvert right movers and left movers, we obtain path interference phenomena which are consistent with a quantum walk.

\begin{figure}[t]
\centering
\includegraphics[width=0.45\textwidth]{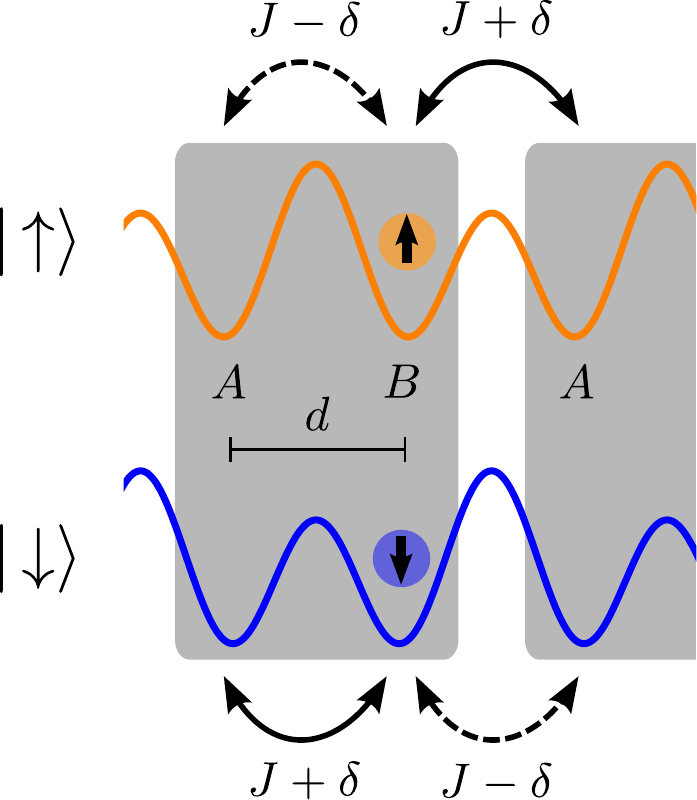}
\caption{1D superlattice used to generate the spatial translation operation of the atomic quantum walk. The grey shaded boxes represent unit cells, in which there exists $A$ and $B$ sublattice sites. This lattice geometry is obtained by superimposing two lasers with wavelengths $\lambda_1$ and $\lambda_2$ such that $\lambda_1=2 \lambda_2$; the resulting intersite distance is $d=\lambda_2/2$.
Spin up ($\uparrow$) particles see the orange lattice (top), which has intra-cell hopping $J-\delta$ (dashed arrow) and inter-cell hopping $J+\delta$ (solid arrow). Spin down ($\downarrow$) particles see the blue lattice (bottom), which is identical to the orange lattice, but shifted by $d$. In the figure, $J$ and $\delta$ are depicted as positive parameters. Particles in this lattice are subject to the Hamiltonian \^H$_S$ Eq.\ (\ref{eq:Ht}).}
\label{fig:spin_dependent_lattice}
\end{figure}

The specific background which is needed to obtain the topological properties we desire is the 1D lattice represented in Fig.\ \ref{fig:spin_dependent_lattice}. We will consider the dynamics of an atom with two internal degrees of freedom, which we can refer to as the particle's spin. Spin up ($\uparrow$) particles see a superlattice with two sites per unit cell, with intra-cell hopping amplitude $J-\delta$ and inter-cell hopping $J+\delta$. This type of lattice is obtained by superimposing two standing waves $l_1$ and $l_2$, with wavelengths $\lambda_1=2 \lambda_2$. The easiest way to obtain $l_2$ is to frequency double $l_1$. The distance between neighbouring sites is $d=\lambda_2/2$, and the size of a full unit cell is $2 d$.

Spin down ($\downarrow$) particles see the same lattice as $\uparrow$ particles, but shifted by $d$. This can be done by making $l_1$ attractive for $\uparrow$ particles but repulsive for $\downarrow$ particles, effectively shifting it by a phase of $\pi$. To do this, set $\lambda_1$ to be the so called anti-magic wavelength of the atom, such that the lattice has an equal and opposite  detuning for $\uparrow$ and $\downarrow$ states.

Interesting candidates to play the role of our $\uparrow$ and $\downarrow$ states are the clock states of either Ytterbium or Strontium atoms. These electronic states have narrow transitions, making them long lived, and are well separated in energy, such that their anti-magic wavelength $\lambda_1$ is readily accessible, while $l_2$ remains at approximately the same amplitude for both species. Additionally, these states can be coupled using Raman beams without significantly heating the system.

We drive this system by periodically applying two laser pulses, which induce Raman transitions between $\uparrow$ and $\downarrow$ states. This is the coin operation \^C$_\theta$ of the atomic quantum walk. The amplitude of the coupling, controlled by the angle $\theta$, is proportional to the intensity of the lasers. Additionally, when the angle between the two lasers is non-zero, \^C$_\theta$ applies a momentum kick.

We will show in Sec.\  \ref{sec:TopoProp} that the topological properties of this system can be modified by changing the value of $\theta$ or $\delta$. We suggest to vary $\theta$ spatially to create a boundary between two regions with different topological properties. One way to do this is to create a gradient in the intensity of the lasers which generate the Raman operation. Because these beams can have beam waists of order the length of the system itself, this can be done simply by focussing the laser away from the centre of the lattice. The amplitude of the spin mixing \^C$_\theta$ is proportional to the intensity of the Raman pulses; thus if the Raman lasers' intensity varies over the length of the system, the spin mixing angle $\theta$ will vary accordingly.

\section{Model}
\label{sec:Model}

In this section, we present the model realizing the atomic quantum
walk, and the operator controlling its time evolution. Using this
knowledge, we perform numerical simulations to determine the system's
properties. Finally, we introduce a convenient unitary transformation
which simplifies the description of the system.

As shown in Fig.\ \ref{fig:spin_dependent_lattice}, we divide the
system into unit cells (indexed by $n$), and assume that the atoms
only ever populate four quantum states per unit cell: within the $n$th
unit cell, the atom can have spin $\uparrow$ or $\downarrow$, and can reside in the
motional ground state of the left/right potential well -- we will
refer to this as sublattice $A/B$. We gather these internal degrees of
freedom into a formal vector, and define vector creation operators as
\begin{equation}
\ccc_n^\dagger = (c^\dagger_{n\uparrow A},c^\dagger_{n\downarrow A}, 
c^\dagger_{n\uparrow B},c^\dagger_{n\downarrow B}).
\end{equation}
We use $\sigma_j$ and $\tau_j$ to denote the Pauli matrices acting in
spin space ($\uparrow$ and $\downarrow$) and sublattice space ($A$ and
$B$) respectively, with $j=\{1, 2, 3\}$, and $\tau_0$ and $\sigma_0$ are $2\times 2$ identity matrices. We will also use $\tau_\pm$ to represent
the sublattice index raising and lowering operators,
\begin{equation}
\label{eq:tau_pm}
\tau_\pm=\frac{1}{2}(\tau_1\pm i \tau_2).
\end{equation}

\subsection*{Shift operation on wavepackets}

Consider first the Hamiltonian of the 1D bichromatic
optical lattice without the Raman pulses.  This reads
\begin{equation}
\begin{split}
\label{eq:Ht}
\text{\^H}_S & =\sum_{n} \ccc_n^\dagger \tau_1 \otimes 
(J \sigma_0 + \delta \sigma_3) \ccc_n\\
& +\sum_n \left( 
\ccc_{n+1}^\dagger \tau_+ \otimes (J \sigma_0-\delta \sigma_3)  \ccc_n + 
\text{H.c}\right).
\end{split}
\end{equation}
The parameters $J$ and $\delta$ control the particles' hopping
amplitudes, as shown in Fig.\ \ref{fig:spin_dependent_lattice}. 
Since the Hamiltonian of Eq.~\eqref{eq:Ht} is translation invariant,
its eigenstates are plane waves with well defined quasimomentum $k\in [-\pi/(2  d), \pi/(2  d)]$. These
states can be chosen to be fully polarised, either $\uparrow$ or $\downarrow$.
Note that each energy eigenvalue is
doubly degenerate, since the system presents two identical lattices
(one for each spin) with no tunnelling between them (see
Fig.\ \ref{fig:spin_dependent_lattice}).

\begin{figure}[t]
\centering
\includegraphics[width=0.45\textwidth]{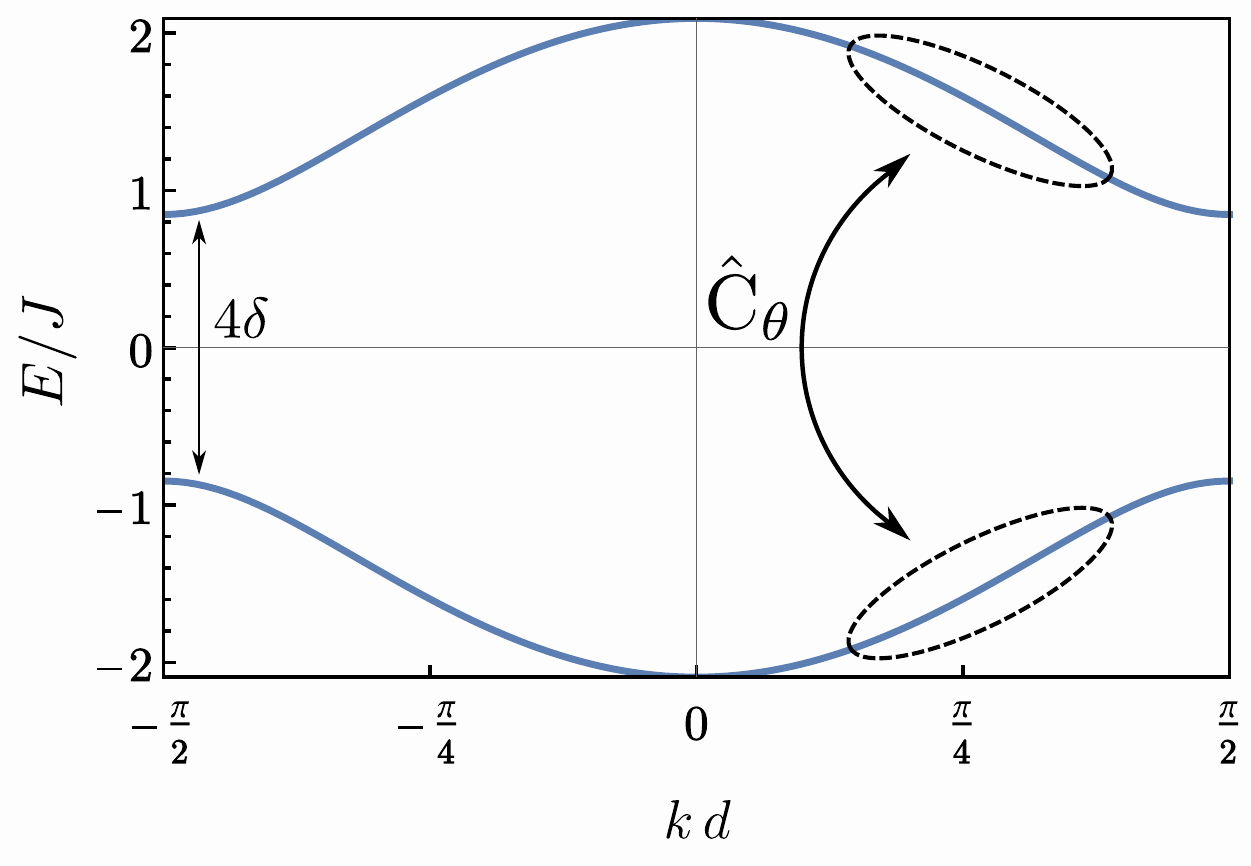}
\caption{Dispersion of \^H$_S$ Eq.\ (\ref{eq:Ht}) for $J=\pi/3$ and $\delta=0.42$. The eigenstates of \^H$_S$ can be chosen to be fully polarised (either $\uparrow$ or $\downarrow$), and the $\uparrow$ and $\downarrow$ bands overlap exactly; the value of the band gap is $4\delta$. Note that the dispersion is symmetric about $E=0$. This means that the slope of the two circled regions of the dispersion are equal and opposite. Thus the wavepackets existing in these regions, which are the states centred around $k=\pi/(4d)$, with energies $E$ and $-E$ respectively, move on average at equal and opposite velocities. The spin mixing \^C$_\theta$ couples states in these regions (illustrated by an arrow).}
\label{fig:Ht_dispersion}
\end{figure}

The object that undergoes a quantum walk is a wavepacket that is broad
in position space but restricted in momentum space to the vicinity of
the wavenumber $k=\pi/(4 d)$.  For this quasimomentum the Hamiltonian
is almost dispersionless, as shown in Fig.\ \ref{fig:Ht_dispersion}.
Thus a wavepacket constructed with states from the lower branches of
the dispersion relation, with $k\approx \pi/(4 d)$, is translated
with a uniform velocity to the right, and its real-space width grows
only very slowly. It is therefore a right-mover. A wavepacket
similarly constructed, but belonging to the upper branches of the
dispersion relation, is a left-mover.

\subsection*{Rotation operation using the Raman pulses}

Consider next the effect of the two Raman lasers on the system, which
are switched on for a brief but intense pulse of duration $\epsilon$. Assuming the laser pulses are intense enough, $\Hs$ can be neglected during the time they are switched on, and we have 
\begin{align}
\label{eq:Htheta}
\text{\^H}(t) &= \sum_n \Omega(n,t) \ccc_{n}^\dagger \tau_3\otimes\sigma_2 \ccc_{n},
\end{align}
where $\Omega(n,t)$ is the Rabi frequency of the lasers
at the position of the $n$th unit cell. By setting the two Raman lasers at an appropriate angle, the pulse applies a $\pi/d$ quasimomentum kick which results in the $\tau_3$ operator appearing in Eq.\ (\ref{eq:Htheta}). We define the area of the pulse at unit cell $n$ by $\theta(n)$:
\begin{equation}
\theta(n) = \int_{-\epsilon/2}^{\epsilon/2} \Omega(n,t) dt,
\end{equation}
such that the effect of the whole pulse is given by:
\begin{align}
\label{eq:C}
\text{\^C}_\theta = \sum_n \ccc_{n}^\dagger 
\exp(-i \theta(n) \tau_3\otimes\sigma_2) \ccc_n \equiv\exp(-i \Htheta).
\end{align}
The effect of $\CCC$ is to couple right-moving states from the bottom branch of the dispersion relation to left-moving states of the top branch, as indicated in Fig.~\ref{fig:Ht_dispersion}. Interestingly, $\CCC$ also couples states on the same branch of the dispersion relation. We will discuss the consequences of this later in the paper.

\subsection*{Complete sequence}

\begin{figure}[t]
\centering
\includegraphics[width=0.49\textwidth]{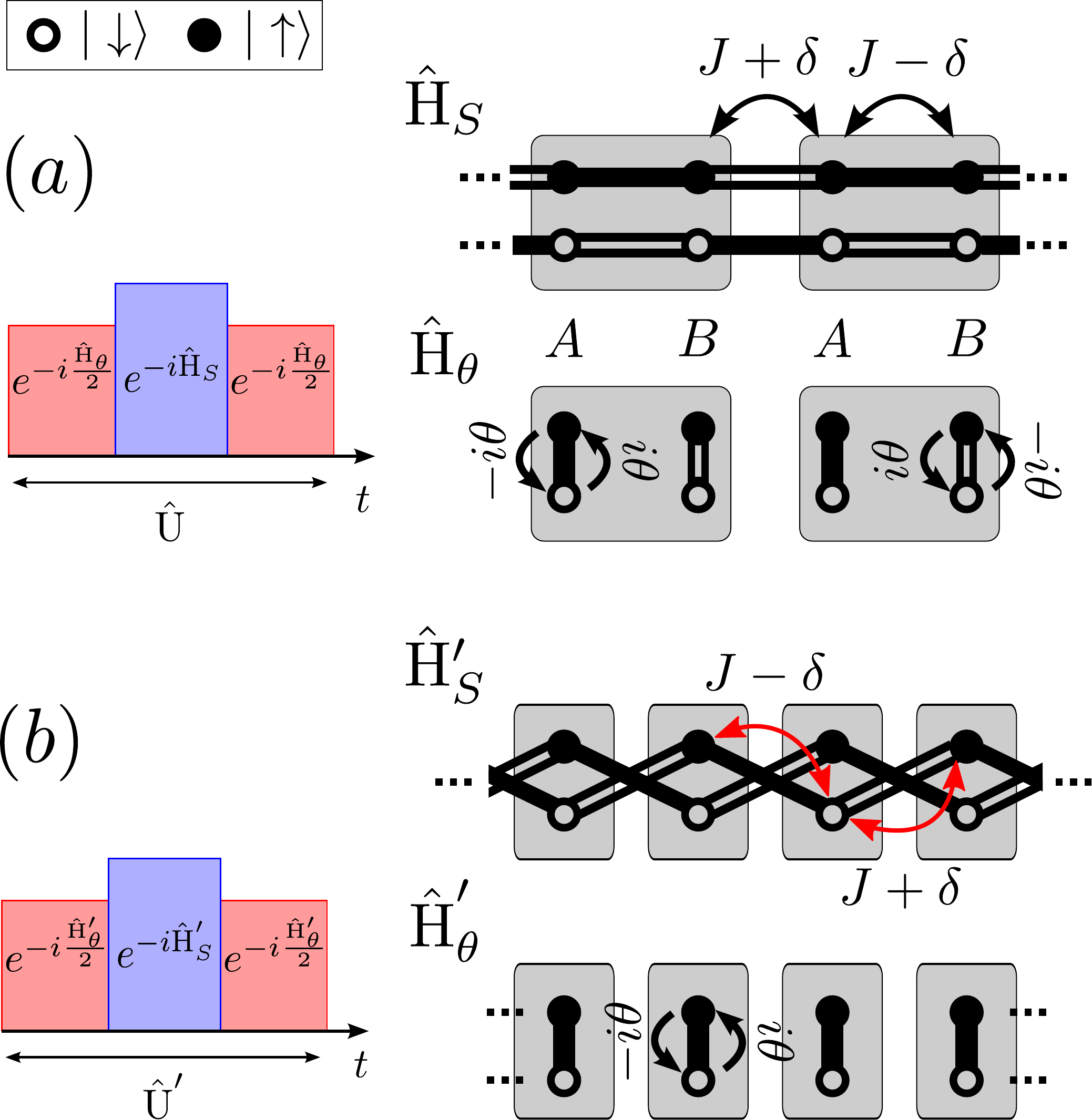}
\caption{Sketch of the tunnelling amplitudes in the atomic quantum walk. Tunnelling amplitudes are different along single and double lines. The tunnelling amplitudes are indicated by curved arrows, the colour of which is unimportant. Unit cells are represented by grey shaded boxes. (a) The time evolution \^U is controlled by \^H$_S$ and \^H$_\theta$, Eqs.\ (\ref{eq:Ht}) and (\ref{eq:C}), respectively. The system has two sites per unit cell. (b) We flip the spins on every second site. the time evolution \^U$'$ is controlled by \^H$_S'$ and \^H$_\theta'$, given by Eqs.\ (\ref{eq:Ht'}) and (\ref{eq:Htheta'}). The lattice has one site per unit cell. The system maps onto the Creutz ladder when $J=\delta$.}
\label{fig:tunnelling_sketches}
\end{figure}

We obtain a quantum walk by switching on the Raman lasers for brief
intense pulses of duration $\epsilon$ which follow each other
periodically, with period $T \gg \epsilon$. In the following, we will
use dimensionless units where $T/\hbar =1$.  The unitary time
evolution operator for one complete period, \^U, reads
\begin{equation}
\label{eq:U}
\text{\^U}=e^{-i\Htheta/2}e^{-i\Hs}e^{-i\Htheta/2}.
\end{equation}
In writing down Eq.\ (\ref{eq:U}), we have chosen the origin of time such that the sequence of operations defining the
walk has an inversion point around which it is symmetric in time, as
discussed in detail in Ref.\ \cite{Asboth2013}.
While this choice has little effect on the system's properties, the form of Eq.\ \eqref{eq:U} makes it easier to find the
symmetries of \^U.
It is however important to always pick the same origin of time when averaging over multiple runs of the experiment, otherwise important details could be averaged out.
The tunnelling amplitudes induced by \^H$_S$ and \^H$_\theta$ are sketched in the figure \ref{fig:tunnelling_sketches}(a).

Finally, we define the Floquet Hamiltonian \^H$_F$ as:
\begin{equation}
\label{eq:HF}
\text{\^U}=\exp(-i \text{\^H}_F).
\end{equation}
This static Hamiltonian describes the motion, integrated over a time
step. This allows us to compute the system's spectrum. Because the
eigenvalues of \^H$_F$ are defined from Eq.\ \eqref{eq:HF}, the band
structure of \^H$_F$ is $2\pi$ periodic. We will refer to the
eigenvalues of \^H$_F$ as the system's quasienergies.

At the end of this section we will find a change of basis which
simplifies \^H$_F$. Despite this, we will prefer to work in the basis
defined above when performing numerical simulations, so that our
results can be easily compared to experimental results from the
protocol described in Sec.\ \ref{sec:Exp}.

\subsection*{Quantum Walk}

\begin{figure}[t]
\centering
\includegraphics[width=0.49\textwidth]{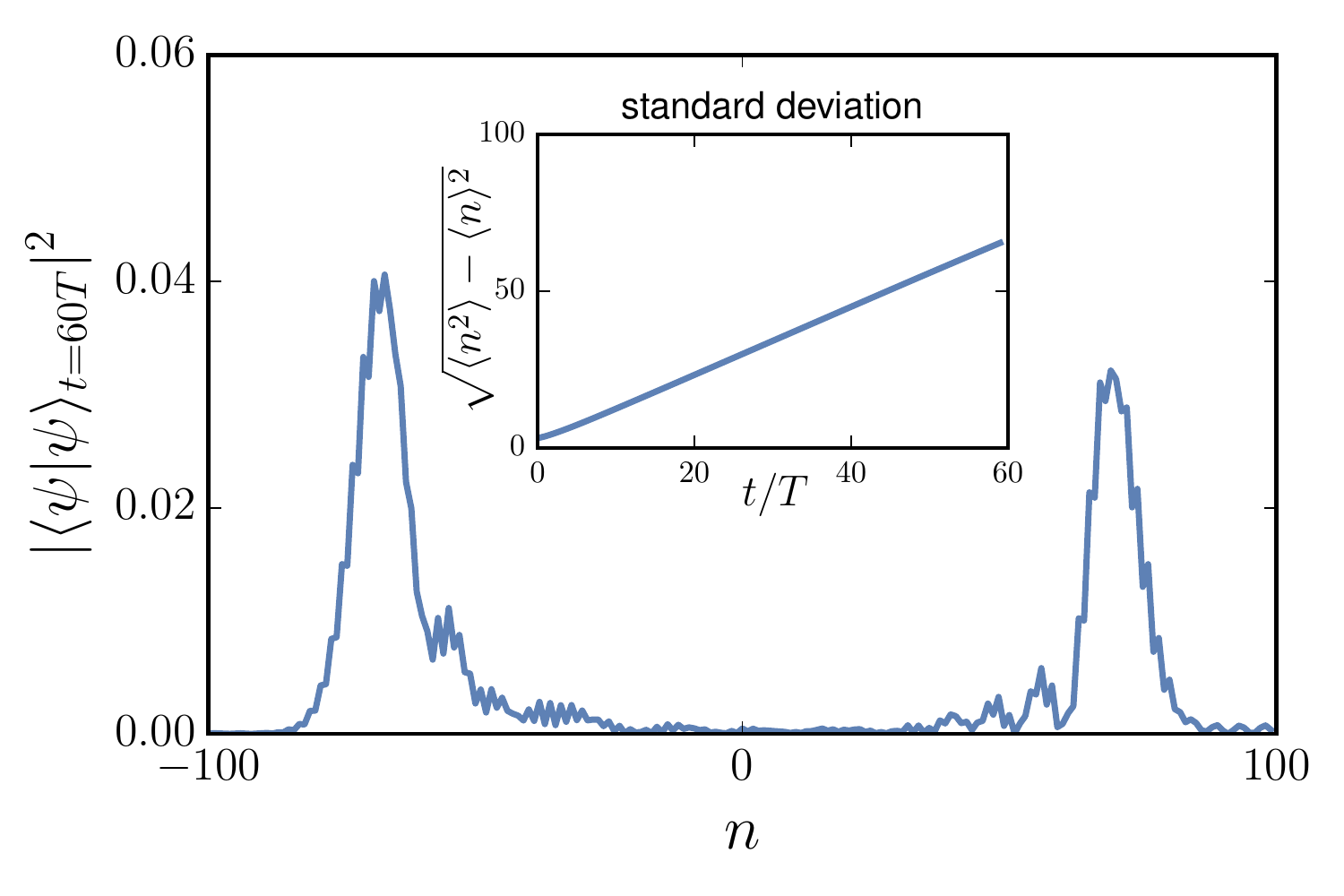}
\caption{Spatial probability distribution after $60$ time-steps. The interference pattern that the density distribution forms is typical of quantum walks. The walker's initial positions were sampled from the Gaussian wave function centred at the site $n=0$, with width $4  d$ and average quasimomentum $k=\pi/(4 d)$. We used the Hamiltonian's parameters: $J=\pi/3$, $\delta=0.42$ and $\theta=0.15$. Inset: the variance of the wavepacket scales linearly with time, which is a characteristic feature of quantum walks.}
\label{fig:AQW_simulation}
\end{figure}

As discussed above, repeated application of the timestep operator
$\UU$ of Eq.\ (\ref{eq:U}) on a wavepacket can be described as a
quantum walk. We verified this using numerical simulations. As an example, we present the atom's final density distribution after $60$ time steps in Fig.\ \ref{fig:AQW_simulation}. For this simulation, we initiated the system with a $\uparrow$ polarised Gaussian wavepacket $|\psi\rangle_{t=0}$ with width $4  d$ centred around $k=\pi/(4 d)$; we set $J=\pi/3$, $\delta=0.42$ and $\theta=0.15$. The final density distribution shows sharp peaks where density is furthest from the origin. The probability to find the particle in any other region is inhibited due to back travelling waves, which interfere destructively with forward travelling ones. The inset shows the standard deviation of the position of the particle, which can be seen to increase linearly with time. Both the destructive interference and the ballistic expansion are well known feature of quantum walks (in contrast with classical random walks, which show diffusive expansion).

\subsection*{Gauge transformation for a smaller unit cell}

By inspection of Fig.\ \ref{fig:tunnelling_sketches}(a), we notice
that it is possible to simplify the system's description by
introducing the new vector creation operators:
\begin{align}
\cc_{2n}^\dagger &= (c^\dagger_{n\uparrow A},c^\dagger_{n\downarrow A});\\
\cc_{2n+1}^\dagger &= (c^\dagger_{n\downarrow B},c^\dagger_{n\uparrow B}).
\end{align}
Notice the inversion of the order of $\uparrow$ and $\downarrow$ on
the odd sites.  In this basis, the tunnellings induced by the atomic
quantum walk are represented in
Fig.\ \ref{fig:tunnelling_sketches}(b). Under this transformation, we
see that the atomic quantum walk is reminiscent of the Creutz ladder \cite{Creutz1999, Bermudez2009, Sticlet2014}, a 1D model which is known to support a non-zero
winding number. We will make this correspondence more obvious in
Appendix \ref{sec:MapToCreutz}.

In the basis introduced above, the Hamiltonians \^H$_S$ and \^H$_\theta$ become:
\begin{eqnarray}
\label{eq:Ht'}
\text{\^H}_S & \rightarrow & \text{\^H}_S'=\sum_n \cc^\dagger_{n+1} (J
 \sigma_1 - i  \delta  \sigma_2) \cc_n +\text{H.c};\\
\label{eq:Htheta'}
\text{\^H}_\theta & \rightarrow & \text{\^H}_\theta'=\sum_n \cc^\dagger_n 
\theta(n)  \sigma_2 \cc_n.
\end{eqnarray}
While \^H$_\theta'$  Eq.\ (\ref{eq:Htheta'}) is already in diagonal form, we can diagonalize \^H$_S'$ by Fourier transforming  Eq.\ (\ref{eq:Ht'}):
\begin{equation}
\label{eq:Ht_Kspace}
\text{\^H}_S'(k) = 2  J \cos(k d) \sigma_1 + 2 \delta \sin(k d) \sigma_2,
\end{equation}
with $k\in [-\pi/d,\pi/d]$. Because of the smaller unit cell, the dispersion has now two non-degenerate branches. It is interesting to redefine our right- and left-walkers in this basis, such that the mapping of the system to a quantum walk can be made more obvious. This is done in Appendix \ref{sec:DoubleQW}, where we show that the system has dynamics which is more complex than the standard quantum walk considered by Refs.\ \cite{Karski2009, Kitagawa2010a, Robens2015}.

By analogy with Eqs.\ \eqref{eq:U} and \eqref{eq:HF}, we can introduce the time evolution operator and the Floquet Hamiltonian in the new basis:
\begin{equation}
\label{eq:U'}
\text{\^U}' = e^{-i \text{\^H}_\theta'/2}\cdot e^{-i \text{\^H}_S'}\cdot e^{-i \text{\^H}_\theta'/2}=\exp(-i \text{\^H}_F').
\end{equation}
In the sections that follow, all analytical results will be obtained in the simplified basis of \^H$_F'$. In the next section, we shall see that the system is topologically non-trivial, and that these topological properties are reflected in \^H$_F'$ .

\section{Topological properties of the atomic quantum walk}
\label{sec:TopoProp}

As we have seen in Eq.\ (\ref{eq:C}), \^C$_\theta$ simultaneously brings the system into a superposition of right movers and left movers, and in a superposition of the spin degree of freedom. This results in spin-orbit coupling terms when the dynamics are averaged over a period of the motion, a fact which is essential for the non-trivial topological properties of the Floquet Hamiltonian to appear. We will see in this section that \^H$_F$ can have non-zero winding numbers, which can be modified by changing the value of $\theta$ and $\delta$.

To understand the topological properties of this system, it is important to analyse the symmetries of \^H$_F'$, the Floquet Hamiltonian in the new basis. The method for finding the symmetries of \^H$_F'$ is explained in detail in Appendix \ref{sec:Symmetries}. By inspection of Eqs.\ (\ref{eq:Htheta'}) and ({\ref{eq:Ht_Kspace}), we find that \^H$_F'$ has chiral symmetry (CS) implemented by the operator $\hat{\Gamma}=\sigma_3$.

The presence of CS implies that \^H$_F'$ anti-commutes with a unitary, Hermitian matrix $\hat{\Gamma}$. Importantly, the CS operator acts only within single unit cells. This will later allow us to break translational invariance without breaking CS. The existence of CS implies that every eigenstate of \^H$_F'$ has a chiral partner with equal and opposite quasienergy:
\begin{equation}
\text{\^H}_F'|\psi\rangle=E|\psi\rangle \Rightarrow \text{\^H}_F'\left(\hat{\Gamma}|\psi\rangle\right)=-E \left(\hat{\Gamma}|\psi\rangle\right).
\end{equation}
States with quasienergies such that $E=-E$ are special because they can transform into themselves under CS. This means that these states can exist without a chiral partner. When this is the case, they cannot be moved away from the quasienergy satisfying $E=-E$ without breaking CS. These states can therefore not be coupled to other states in the system by chiral symmetric perturbations.

In a Floquet system the quasienergies are $2\pi$-periodic. There are therefore two quasienergies, $E=0$ and $E=\pi$, which satisfy the condition $E=-E$. As we will show shortly, this implies that bound states which are topologically protected as a result of CS can appear at $E=0$ or $E=\pi$.

A system which presents CS but neither time reversal symmetry nor particle hole symmetry belongs to the AIII class of the topological classification of Hamiltonians \cite{Schnyder2008}. Floquet Hamiltonians in 1D which belong to this class can have two non-zero topological invariants, the winding numbers $\nu_0$ and $\nu_\pi$ \cite{Asboth2013}. For a detailed physical interpretation of the physical significance of these quantities, we refer the reader to Ref.\ \cite{Asboth2014}.

Consider connecting two regions $R_1, R_2$ which have different winding numbers and present spectral gaps. The topological invariants $\nu_0,\nu_\pi$ cannot change without closing the spectral gap in $E=0, E=\pi$ respectively. We assume that the boundary between the two bulks is smooth and slowly varying, such that the spectral gaps of $R_1, R_2$ are densely populated in this region. If we now make the interface between $R_1$ and $R_2$ sharper, fewer states can live in the boundary region, meaning that fewer states can populate the spectral gap of $R_1, R_2$. To illustrate this, the energy of the three states closest to $E=0$ at a topological boundary are  plotted versus the boundary sharpness in Fig.\ \ref{fig:SSH_first_excited_state}. In the infinitely sharp boundary limit, all states have been lifted out of the spectral gap except for those which do not have a chiral partner. Thus, if a system presents an interface between regions $R_1, R_2$ with different winding numbers $\nu_0$, there must exist an $E=0$ state at this interface. Because $R_1$ and $R_2$ must be gapped to have well defined winding number, the $E=0$ state can only exist at the topological interface. Correspondingly, if the system presents an interface between two values of $\nu_\pi$, an $E=\pi$ bound state must exist where $\nu_\pi$ changes.

\begin{figure}[t]
\centering
\includegraphics[width=0.45\textwidth]{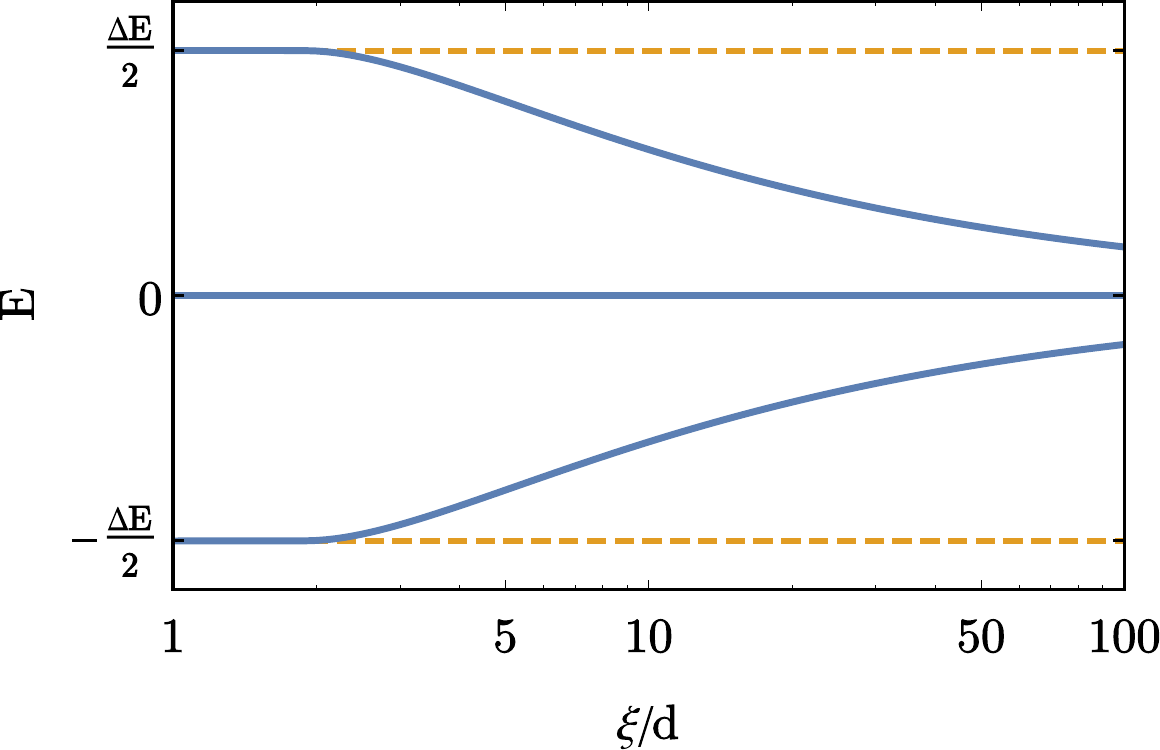}
\caption{Energy of the three states closest to $E=0$ at a topological boundary between two regions $R_1,R_2$ versus the boundary sharpness, controlled by the boundary width $\xi$. The system is chiral symmetric and regions $R_1,R_2$ have a spectral gap $[-\Delta E/2,\Delta E/2]$; these energies are indicated by dashed yellow lines. As $\xi$ is reduced, the interface between $R_1$ and $R_2$ can accommodate fewer states, such that, when $\xi\lesssim 1$, only topological bound states can exist in the spectral gap of $R_1,R_2$. }
\label{fig:SSH_first_excited_state}
\end{figure}

In fact, the number of times that the spectral gap closes at the interface between two regions is (at least) the difference in their winding numbers. The difference in $\nu_0$ ($\nu_\pi$) tells us how many topologically protected $E=0$ ($E=\pi$) bound states exist at this interface. In the following, we will follow Refs.\ \cite{Asboth2013,Asboth2014} to calculate the winding numbers $\nu_0$ and $\nu_\pi$ from the time evolution \^U$'$, and find the parameter regime which makes the atomic quantum walk topologically non-trivial.

The time evolution operator Eq.\ (\ref{eq:U'}) has the form:
\begin{equation}
\text{\^U}'=\hat{\Gamma}\cdot \text{\^G}^\dagger\cdot \hat{\Gamma} \cdot \text{\^G},
\end{equation}
with
\begin{equation}
\label{eq:F}
\text{\^G}=e^{-i \text{\^H}_S'/2}\cdot e^{-i \text{\^H}_\theta'/2}
=\begin{pmatrix}
a(k) & b(k)\\
c(k) & d(k)
\end{pmatrix},
\end{equation}
and where $a(k), ...,  d(k)$ are the entries of \^G, which are, in general, complex functions of $k$. In general, the winding number of a function $z(k)$ can be evaluated using the formula:
\begin{equation}
\label{eq:evaluate_winding}
\nu[z]=\frac{1}{2\pi i}\int_{-\pi/d}^{\pi/d}dk \frac{d}{dk}\log z(k).
\end{equation}
In the basis where $\hat{\Gamma}=\sigma_3$, we have $\nu_0=\nu[b]$ and $\nu_\pi=\nu[d]$, with $\nu[b]$ and $\nu[d]$ the winding numbers of the $b(k)$ and $d(k)$ functions respectively \cite{Asboth2013}.

Due to the anti-commutation property of Pauli matrices, we have that for any vector $\bm{v}$:
\begin{equation}
\label{eq:SigmaAntiCommut}
e^{i \bm{v}\cdot\bm{\sigma}}=\sigma_0 \cos(|\bm{v}|)+i \frac{\bm{v}\cdot \bm{\sigma}}{|\bm{v}|} \sin(|\bm{v}|).
\end{equation}
This allows us to compute the exponential forms of Eqs.\ (\ref{eq:Htheta'}) and (\ref{eq:Ht_Kspace}). Substituting into Eq.\ (\ref{eq:F}), we find:
\begin{equation}
\label{eq:b(k)}
\begin{split}
b(k) &= \cos(\varepsilon(k)) \sin(\theta/2)\\
&-\sin(\varepsilon(k)) \cos(\theta/2) \frac{\delta \sin(k d)+i   J \cos(k d)}{\varepsilon(k)},
\end{split}
\end{equation}
and:
\begin{equation}
\label{eq:d(k)}
\begin{split}
d(k) &= \cos(\varepsilon(k)) \cos(\theta/2)\\
&+\sin(\varepsilon(k)) \sin(\theta/2) \frac{\delta \sin(k d)-i  J \cos(k d)}{\varepsilon(k)},
\end{split}
\end{equation}
where $\varepsilon(k)$ are the energy eigenvalues of \^H$_T'/2$:
\begin{equation}
\varepsilon(k)=\pm\sqrt{J^2 \cos^2(k d)+\delta^2 \sin^2(k d)}.
\end{equation}

\begin{figure}[t]
\centering
\includegraphics[width=0.49\textwidth]{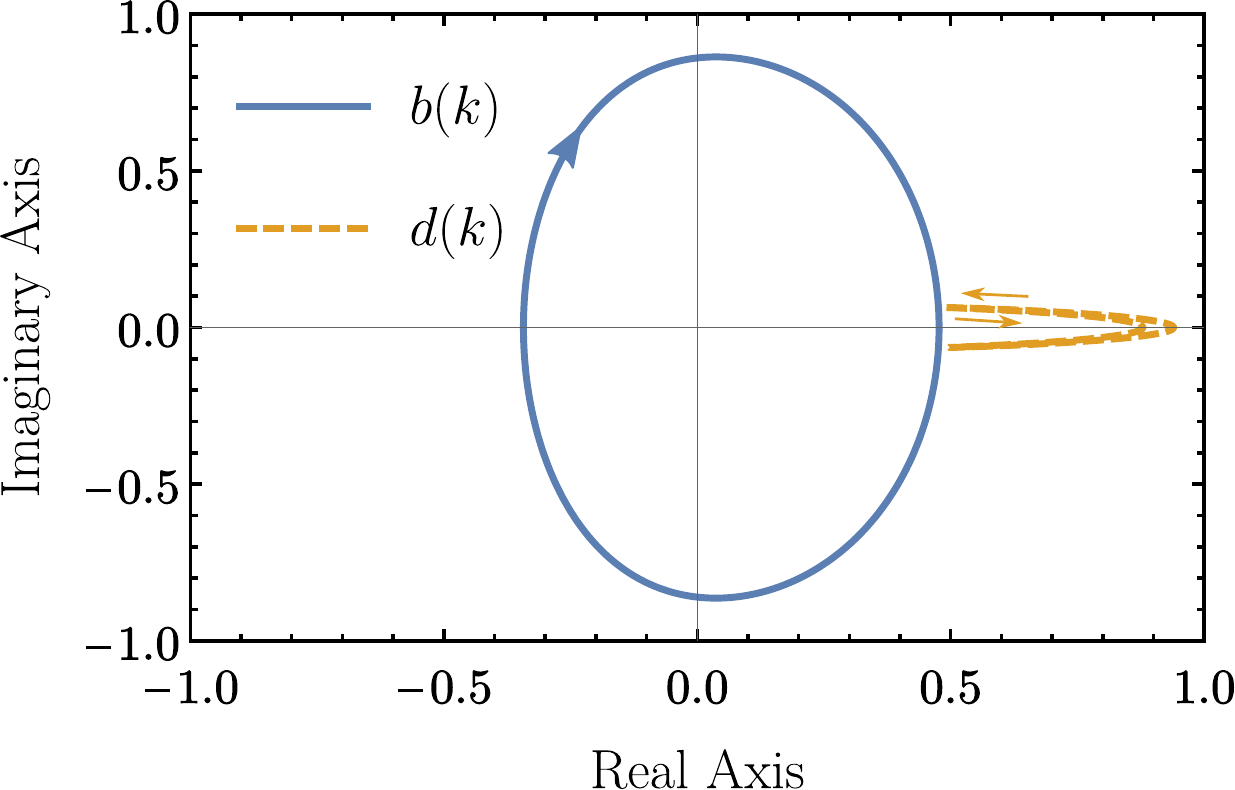}
\caption{Plots of functions $b(k)$ Eq.\ (\ref{eq:b(k)}) (full blue) and $d(k)$ Eq.\ (\ref{eq:d(k)}) (dashed yellow) in the complex plane as $k$ goes from $-\pi/d$ to $\pi/d$ for $J=\pi/3$, $\delta=0.42$, and $\theta=0.15$. The number of times that $b(k)$ ($d(k)$) winds around the origin corresponds to the topological invariant $\nu_0$ ($\nu_\pi$). In this case we can read off $\nu_0=-1$ and $\nu_\pi=0$.}
\label{fig:winding_example}
\end{figure}

The simplest way to visualize the winding numbers $\nu_0$ and $\nu_\pi$ is to plot $b(k)$ and $d(k)$ in the complex plane as $k$ goes from $-\pi/d$ to $\pi/d$ (both bands have the same winding number). The system is topologically non-trivial if at least one of the curves winds around the origin. The curves that $b(k)$ and $d(k)$ form in the complex plane are presented for $J=\pi/3$, $\delta=0.42$, and $\theta=0.15$ in Fig.\ \ref{fig:winding_example}. It is clear from this figure that the system can present non-zero winding numbers.

The winding number $\nu_0$ cannot change unless both the real and imaginary parts of $b(k)$ vanish simultaneously (see Fig.\ \ref{fig:winding_example}). This happens when $\delta=\pm\theta/2+n \pi,~n\in\mathbb{Z}$, for which values the band gap closes either at quasimomentum $k=\pi/(2 d)$ or $k=-\pi/(2 d)$. Similarly, $\nu_\pi$ cannot change unless $d(k)$ vanishes, which happens when $\delta=\pm\theta/2+(n+1/2) \pi,~n\in\mathbb{Z}$, at $k=\pi/(2 d)$ or $k=-\pi/(2 d)$. This allows us to construct the topological phase diagram of the atomic quantum walk, which is presented in Fig.\ \ref{fig:phase_diagram}. We evaluate the winding numbers in each region using the general formula Eq.\ (\ref{eq:evaluate_winding}). It is also possible for band gap closing events to occur in $\theta=n\pi,~n\in\mathbb{Z}$ without changing either of the winding numbers. When $J=\pi/3$ however (value used in the simulations presented in this paper), no such events occur in the interval considered in the Fig.\ \ref{fig:phase_diagram}.

\begin{figure}[t]
\centering
\includegraphics[width=0.49\textwidth]{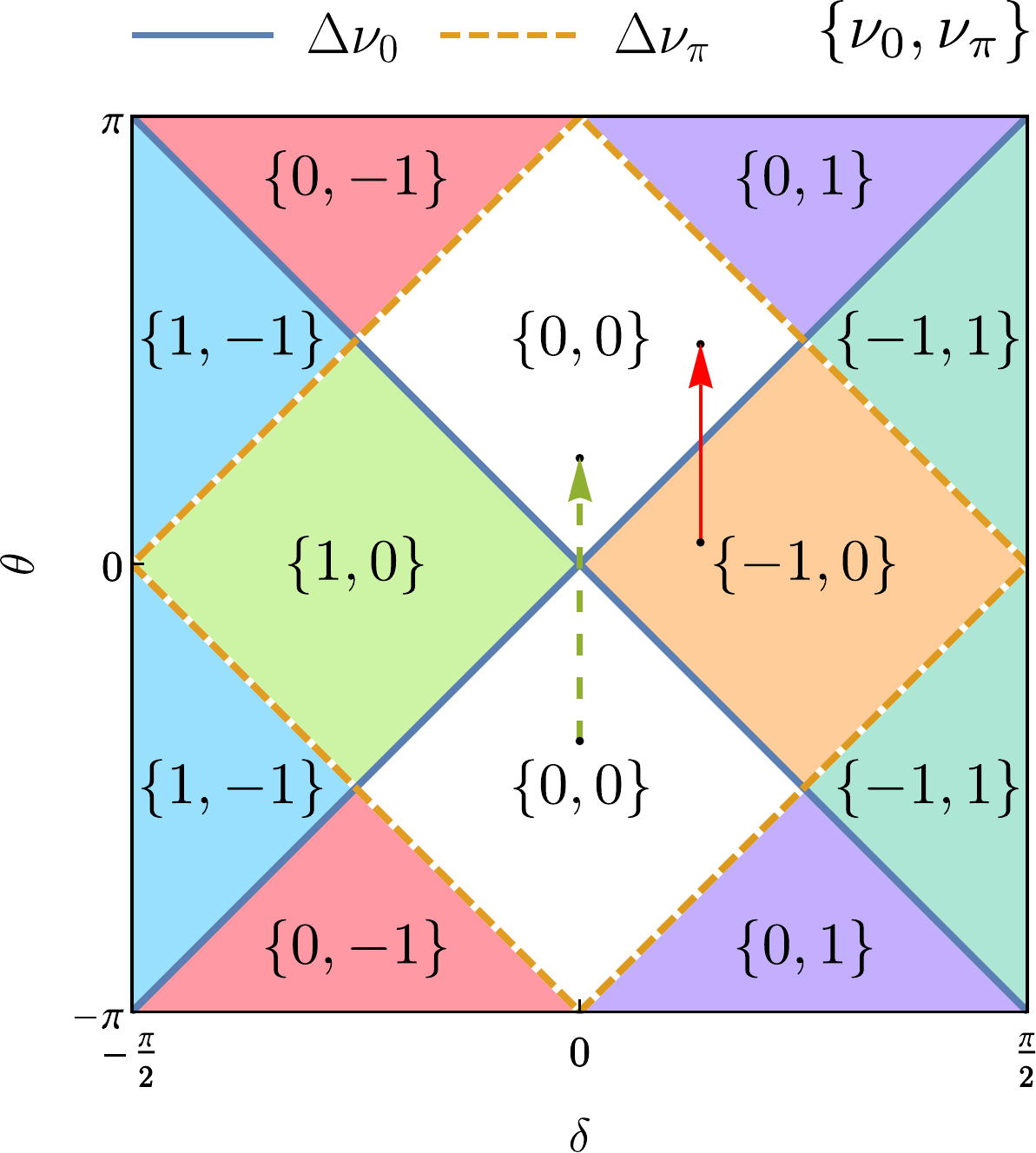}
\caption{Phase diagram of the atomic quantum walk. The function $b(k)$ ($d(k)$) vanishes along the full blue (dashed yellow) lines, indicating a boundary along which the topological invariant $\nu_0$ ($\nu_\pi$) can change its value. The winding numbers for each region of the plot have been calculated using Eq.\ (\ref{eq:evaluate_winding}), and are specified on the figure as $\{\nu_0,\nu_\pi\}$; similarly coloured regions have the same winding numbers. In Sec.\ \ref{sec:BoundState}, we produce an $E=0$ bound state by varying $\theta$ spatially along the path in parameter space indicated by a red arrow. The topological boundary $\theta=2 \delta$ is crossed in this process. In Sec.\ \ref{sec:PairOfJR}, we cross a trivial band gap closing which occurs for $\delta=0$. This path in parameter space is indicated by a dashed green arrow.}
\label{fig:phase_diagram}
\end{figure}

As is visible from Fig.\ \ref{fig:phase_diagram}, the winding number of the atomic quantum walk is a function of $\delta$ and $\theta$. In the following section, we will use this to create two regions with distinct topological properties.

\section{Detection of the topological bound state}
\label{sec:BoundState}

As we saw in Sec.\ \ref{sec:TopoProp}, the topological properties of the atomic quantum walk can be modified by changing the spin mixing angle $\theta$. By using spatially inhomogeneous Raman lasers, it is therefore possible to create two regions in the system with distinct topological properties. At the boundary between two regions which have different winding numbers, there lives a robust bound state which is protected by the system's symmetries and pinned either at $E=0$ or $E=\pi$. In the following, we suggest a method for experimentally generating this topological bound state, and identifying it using its characteristic spin distribution. The density distribution can either be retrieved in a single measurement, if the quantum walk is performed with a gas of non-interacting atoms, or by repeating the experiment many times and averaging if a single atom is used. We will be interested in the spin populations of sublattices $A$ and $B$, meaning that the position measurement must have single site resolution and be sensitive to the atom's spin. The location of $A$ and $B$ sites is fixed by the lasers generating the optical lattice (see Fig.\ \ref{fig:spin_dependent_lattice}), so that their position will remain the same from one experiment to the next.

\begin{figure}[t]
\centering
\includegraphics[width=0.49\textwidth]{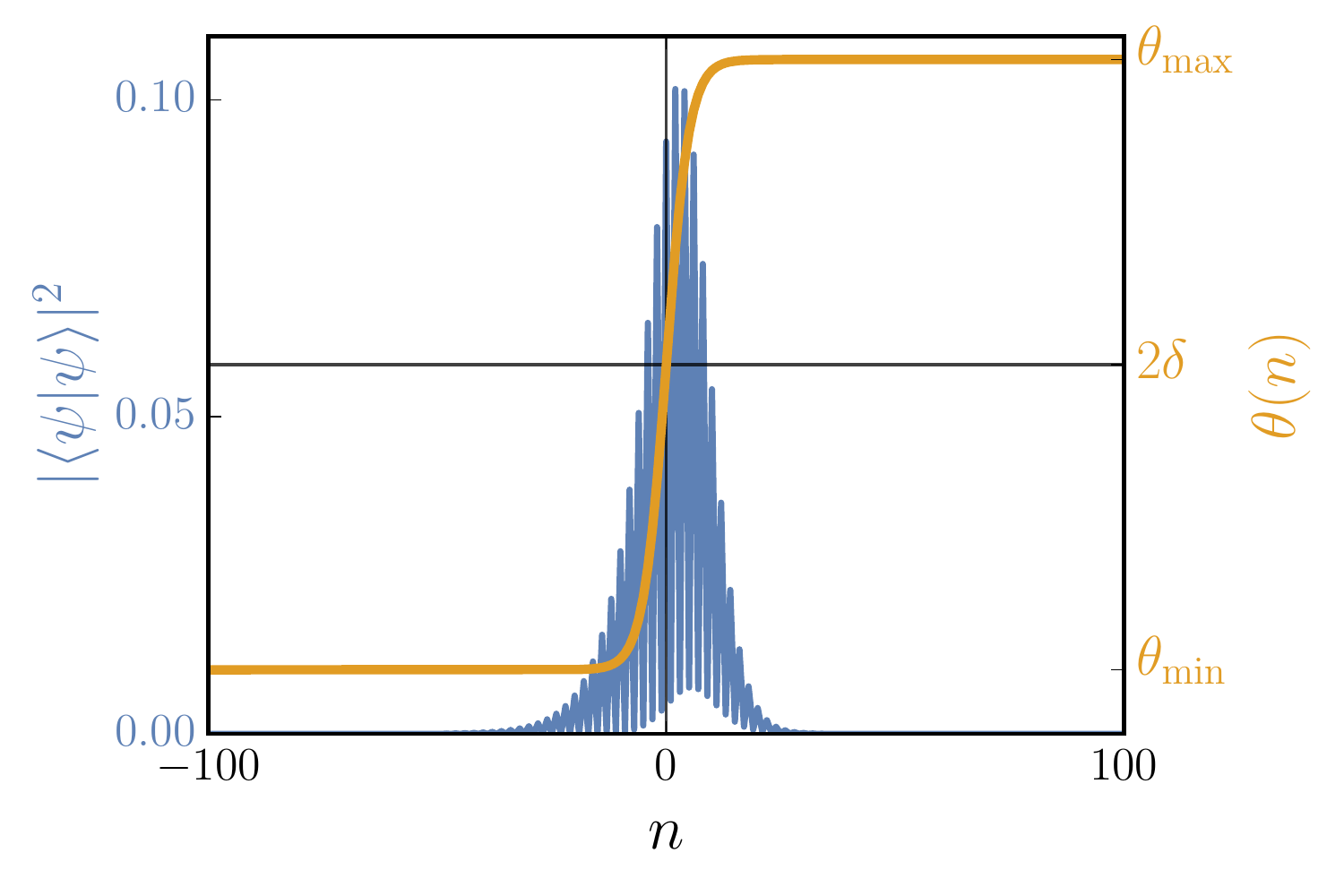}
\caption{Topological interface created by varying $\theta$ spatially from $\theta_{\rm min}$ to $\theta_{\rm max}$. In yellow: $\theta$ varies spatially according to Eq.\ (\ref{eq:theta_function}). The topological boundary occurs at $n=0$, which is the point where $\theta=2 \delta$. In blue: topological bound state occurring at this boundary, obtained by exact diagonalisation of \^H$_F$ for $J=\pi/3$, $\delta=0.42$, $\theta_{\rm min}=0.15$, $\theta_{\rm max}=1.5$ and $\xi=10d$. This path in parameter space is indicated on Fig.\ \ref{fig:phase_diagram} by a red arrow.}
\label{fig:theta_function}
\end{figure}

When performing this experiment, we expect a portion of the density to remain pinned to the topological boundary. This corresponds to the part of the initial state which overlaps with the bound state wave function. The rest of the density is translated ballistically away from the topological interface. We verify this numerically by performing simulations, where $\theta$ is varied according to:
\begin{equation}
\label{eq:theta_function}
\theta(n)=\frac{\theta_{\rm max}+\theta_{\rm min}}{2}+\frac{\theta_{\rm max}-\theta_{\rm min}}{2} \tanh\left(\frac{n d}{\xi} \right),
\end{equation}
where $\xi$ determines the width of the region where $\theta(n)$ changes value. This function is represented in Fig.\ \ref{fig:theta_function}. Note that the existence of the bound state only requires that $\theta(n)$ crosses $2\delta$, and does not depend otherwise on the precise form of Eq.\ \eqref{eq:theta_function}. Our simulations are performed in the basis where \^H$_S$ and \^H$_\theta$ have the form given by Eqs.\ (\ref{eq:Ht}) and (\ref{eq:C}) respectively.

\begin{figure*}[t]
\centering
\includegraphics[width=0.99\textwidth]{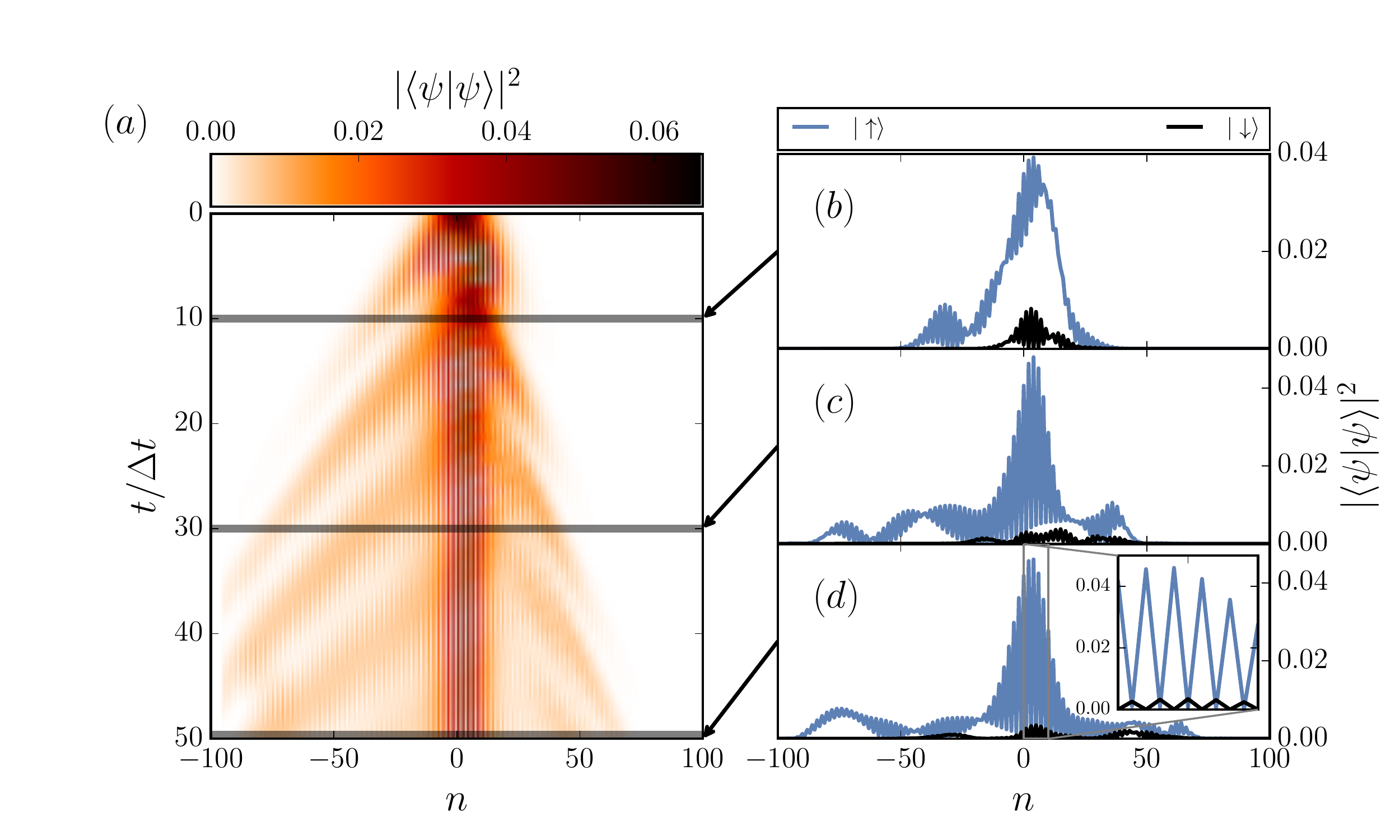}
\caption{(a) Atom density in position space versus time. After $t/T=50$ time-steps, some density has escaped ballistically to infinity but the probability density function remains sharply peaked around the origin. (b), (c) and (d) Probability density function at times $t_1=10$, $t_2=30$ and $t_3=50$ respectively, with $\uparrow$ represented in blue (grey) and $\downarrow$ in black. The inset of (d) shows the density in the interval $n\in [0,  10]$. We observe that $\uparrow$ ($\downarrow$) states have non-zero density only on even (odd) sites. This simulation was realised with $J=\pi/3$, $\delta=0.42$, and $\theta$ varying spatially from $\theta_{\rm min}=0.15$ to $\theta_{\rm max}=1.5$. The initial state was a Gaussian wavepacket centred around site $n=0$ with mean quasimomentum $k=\pi/(2d)$, in $\uparrow$ state with equal support on $A$ and $B$ sublattices.}
\label{fig:pdf_BS}
\end{figure*}

Fig.\ \ref{fig:pdf_BS}(a) shows an example of the evolution of the atomic density during 50 time-steps. As expected, we find that a portion of the density remains pinned to the region of $n=0$, which is the location of the topological boundary. Atoms which do not populate the bound state are transported ballistically away from $n=0$. The spreading is anisotropic due to the initial state we chose, which is fully $\uparrow$ polarised. The spreading density shows interference patterns between forward and backward travelling atoms, as is characteristic for quantum walks (indeed, we already observed this behaviour in Fig.\ \ref{fig:AQW_simulation}).

The density at specific moments in time is represented in Fig.\ \ref{fig:pdf_BS}(b), (c) and (d). We observe that at late times, the probability distribution is exponentially peaked at the location of the topological boundary. Importantly, the relative population of this state does not decrease at later times. We have verified this numerically by computing the overlap between the atom's wave function and the $E=0$ eigenstate of \^H$_F$, and found that it is time independent.

Experimentally, this can be verified by plotting the total density in the neighbourhood of the topological boundary, which is displayed in Fig.\ \ref{fig:FullDensBS}. We see that at early times, the total density near the origin decreases rapidly as atoms which do not populate the bound state are transported away from the topological boundary. During this period, we notice that the total density presents oscillations; these are due to the interference of ballistically transported atoms. At late times, the total density near the topological boundary converges to a non-zero value, which is a sign that atoms populating the bound state do not leak into other states of the system.

\begin{figure}[t]
\centering
\includegraphics[width=0.49\textwidth]{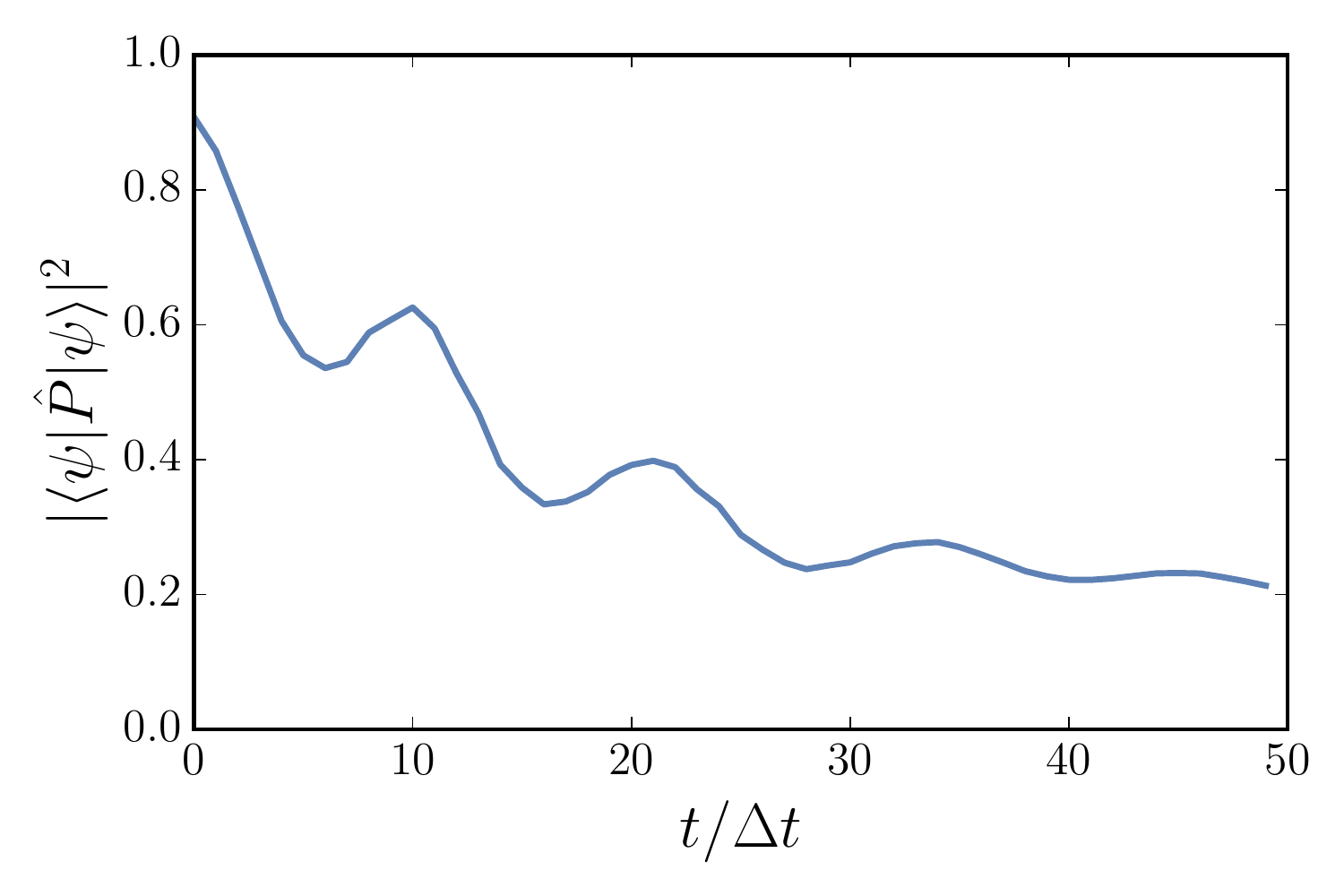}
\caption{Full atomic density, obtained from the simulation Fig.\ \ref{fig:pdf_BS}, for $50$ time-steps in the region $n\in[-15, 15]$. This interval corresponds to the neighbourhood of the topological boundary, and $\hat{P}$ is the projector onto this region. At early times we observe oscillations as atoms which are not trapped in the bound state leave the region of the boundary. At late times, we see that the total density converges to a non-zero value.}
\label{fig:FullDensBS}
\end{figure}

Another experimental signature is the mapping of the phase boundaries. Indeed, Fig.\ \ref{fig:phase_diagram} show the parameter regimes in which we expect to find topological bound states, and knowledge of the function with which $\theta$ varies determines the position around which these states are centred. By varying $\delta$ and $\theta$, we can explore the system's parameter space and verify that these exponentially bound states occur at all topological boundaries. This provides a straightforward way to verify that we are indeed observing topological bound states, and relies only on imaging the atoms' probability density function.

An alternative method for verifying that a system is a topological bound state is to verify that it is an eigenstate of the chiral symmetry (CS) operator $\hat{\Gamma}$. Indeed, we saw in Sec.\ \ref{sec:TopoProp} that if a state verifies the following conditions, it is a topological bound state:
\begin{enumerate}
\item The state is an eigenstate simultaneously of $\hat{\Gamma}$ and the Floquet Hamiltonian \^H$_F$.
\item The state has vanishing overlap with any other state of equal energy which satisfies condition 1.
\end{enumerate}
In the following, we describe a method to identify an eigenstate of $\hat{\Gamma}$, thereby providing a strong way of identifying a topological bound state.

We can verify that the bound state we are observing is a single eigenstate of \^H$_F$ directly from the time evolution of probability density distribution. Indeed, if this state was a superposition of states, we would observe Rabi oscillations, while it is clear from figure Fig.\ \ref{fig:pdf_BS}(a) that the density distribution near the origin remains the same after a period of driving. Additionally, it can be seen from the inset of Fig.\ \ref{fig:pdf_BS}(d) that the state which is found near the origin at late times has an extremely interesting structure. Indeed, we find that $\uparrow$ states only occur on even sites, while $\downarrow$ states only occur on odd sites. This is a direct result of being an eigenstate of CS, which has the form $\hat{\Gamma}=\tau_3\otimes\sigma_3$ in this basis, leading to a constrained spin distribution. Importantly, all other eigenstates of \^H$_F$ must transform into their chiral partner under the action of CS. The only states that can do this have equal $\uparrow$ and $\downarrow$ density on each site. Consequently, the state we are measuring near $n=0$ can only be a topological bound state.

A spin sensitive density measurement therefore provides a direct
method to identify an $E=0$ or an $E=\pi$ bound
state protected by CS in a 1D system. Using the exact same measurement
at half time-steps, it is possible to discriminate between $E=0$ and
$E=\pi$ energy states. A half time-step is performed by applying a
spin rotation $\theta/2$, followed by time evolving with \^H$_S$ for a
time $T/2$. It was shown in \cite{Asboth2014} that the dynamics
of the bound state between two time-steps is sensitive to the state's
energy. As shown in Fig.\ \ref{fig:eigenstate_half_time_step}(a), at
integer time-steps, the $\uparrow$ ($\downarrow$) density of $E=\pi$
states only has support on even (odd) sites. At half time-steps, the
spin structure of $E=\pi$ states is reversed, with spin $\uparrow$
($\downarrow$) states only on odd (even) sites. This is shown in
Fig.\ \ref{fig:eigenstate_half_time_step}(b). This contrasts with the
behaviour of $E=0$ states, which keep the same spin structure at
integer time-steps and half integer time-steps. Thus, performing a
position measurement with single site resolution which is sensitive to
the spin state does not only identify a topological
bound state of the system, but it also provides a method to
differentiate an $E=0$ from an $E=\pi$ state.

\begin{figure}[t]
\centering
\includegraphics[width=0.49\textwidth]{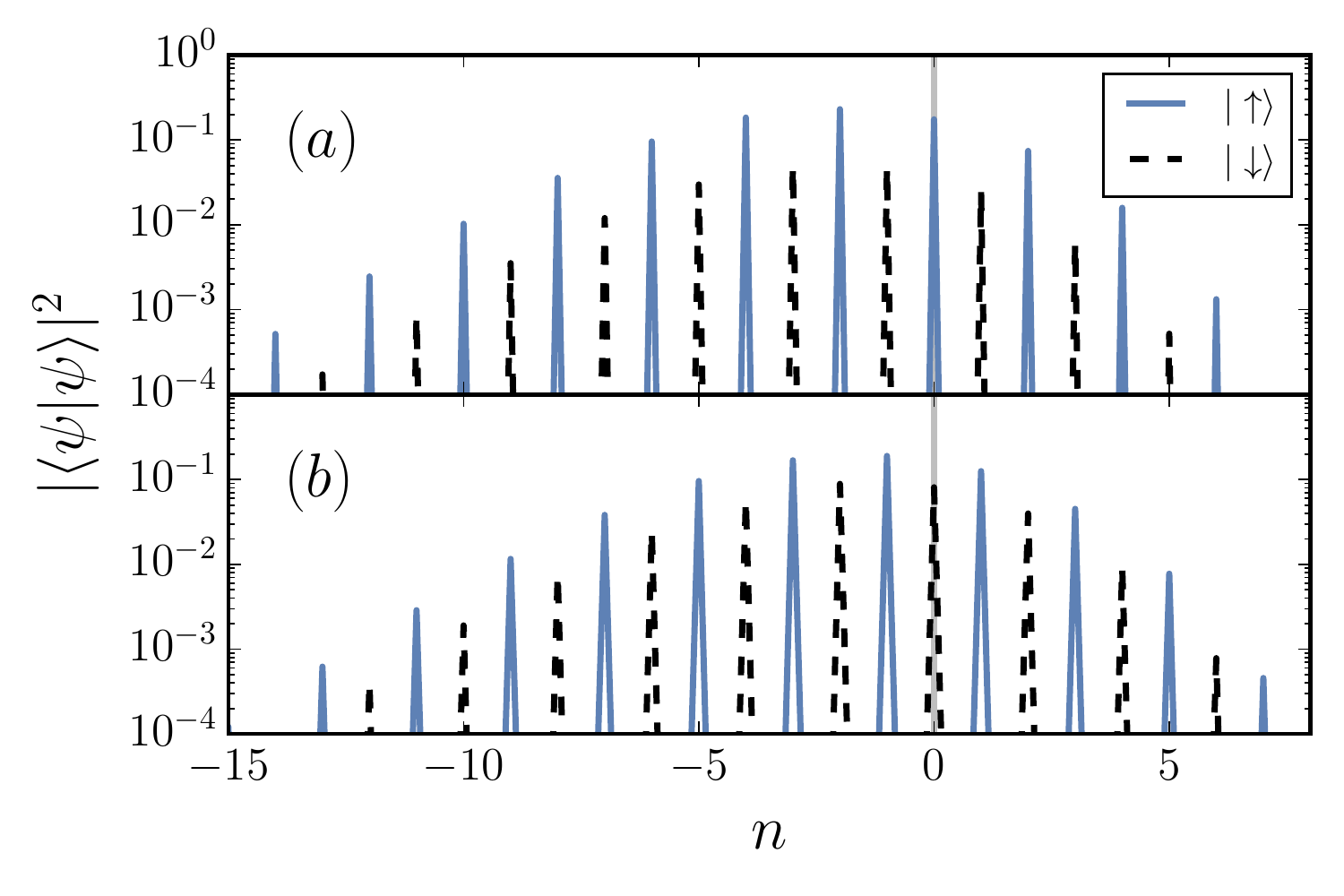}
\caption{(a) Density distribution of an $E=\pi$ topological bound state (plotted on a log scale), obtained by diagonalising \^H$_F$ (not \^H$_F'$). $\uparrow$ ($\downarrow$) density is represented in blue (black). The state is exponentially localised in the neighbourhood of $n=0$. $\uparrow$ ($\downarrow$) states are only found on even (odd) sites. Only eigenstates of the chiral symmetry operator have this  complex spin distribution. (b) The same state, after evolution through half a time-step. $\uparrow$ ($\downarrow$) states are only found on odd (even) sites. For clarity, a grey vertical line has been drawn at $n=0$. States with $E=\pi$ show an inversion in their spin density distribution at half time-steps.}
\label{fig:eigenstate_half_time_step}
\end{figure}

To recapitulate, we suggested a method to generate topologically protected bound states experimentally, and simulated this protocol numerically. We saw that it is possible to identify topological bound states by correlating their occurrence with Fig.\ \ref{fig:phase_diagram} and mapping out the phase boundaries by changing $\theta, \delta$. Alternatively, we can recognise topological bound states thanks to their unique density distribution, which can also be used to differentiate $E=0$ from $E=\pi$ states. In the next section, we will see that exponentially localised states with $E\approx 0, \pi$ can also appear at trivial band gap closings (ones where the winding number does not change). It is therefore important to know for which values of $\delta, \theta$ these occur when exploring the parameter space of the atomic quantum walk.

\section{Pair of momentum separated Jackiw-Rebbi states}
\label{sec:PairOfJR}

In this section, we will see that not all $E= 0, \pi$ states which are exponentially localised are topologically protected. We will consider the atomic quantum walk in the limit $\delta=0$. As can be seen from Fig.\ \ref{fig:phase_diagram}, when following the path in parameter space indicated by a dashed green arrow, a topologically trivial band gap closing occurs when $\theta$ change sign. Indeed, the winding numbers $\{\nu_0,\nu_\pi\}$ are the same in both regions. As a result of this, the Hamiltonian in this limit also presents $E\approx 0$ and $E\approx \pi$ bound states. By finding an approximate expression for the Floquet Hamiltonian \^H$_F'$, we will understand that the $E\approx 0$ bound states correspond to solutions of the Jackiw-Rebbi model in the continuum case. Because these states occur in pairs, they can hybridise and move symmetrically away from $E=0$, thus destroying the states' topological protection. Importantly, these trivial bound states are not eigenstates of chiral symmetry, and therefore do not have the same spin distribution as the states presented in Fig.\ \ref{fig:eigenstate_half_time_step}.

In the limit of $\delta=0$, \^H$_S'$ from Eq.\ \eqref{eq:Ht'} and \^H$_\theta'$ from Eq.\ \eqref{eq:Htheta'} have the form:
\begin{eqnarray}
\label{eq:Ht_SQW}
\text{\^H}_S' &=& \sum_n J \cc^\dagger_{n+1}  \sigma_1  \cc_n +\text{H.c};\\
\label{eq:C_SQW}
\text{\^H}_\theta'&=&-\sum_n \cc^\dagger_{n} \theta(n)  \sigma_2\cc_n .
\end{eqnarray}
Using the method presented in Appendix \ref{sec:Symmetries}, we find the symmetries of the system in this limit. These are presented in the Table \ref{tab:SQW_Sym}. From the expression of the chiral symmetry, time reversal symmetry and particle hole symmetry operators, we find that we are in the BdG symmetry class CI of the classification of single-particle Hamiltonians. Thus, the atomic quantum walk is topologically trivial in 1D in the limit $\delta=0$ \cite{Schnyder2008}.

\begin{table}[t]
\centering
\caption{In the first column, the operators implementing various symmetries of \^H$_F'$ are listed in the limit $\delta=0$. We list the squares of these operators in the second column.}
\label{tab:SQW_Sym}
\begin{tabular}{ |c|c|c| }
\hline
symmetry & operator & square\\
\hline
Chiral symmetry & $\sigma_3$ & $\sigma_0$ \\
Time reversal symmetry & $\sigma_1  \hat{\mathcal{K}}$ & $\sigma_0$ \\
Particle hole symmetry & $-i \sigma_2 \hat{\mathcal{K}}$ & $-\sigma_0$ \\
\hline
\end{tabular}
\end{table}

Despite this, the system displays $E\approx 0$ and $E\approx \pi$ bound states when the spin mixing angle $\theta$ is varied spatially. We can verify this numerically by considering a chain of length $L$ and $\theta=\theta(n)$, a continuous function of position, going from $\theta_{\rm min}$ to $\theta_{\rm max}$ over a length scale $\xi$ according to Eq.\ (\ref{eq:theta_function}). We assume $d \ll \xi \ll L$, where $d$ is the lattice spacing. The spectrum of \^H$_F'$ from Eq. \eqref{eq:U'} is shown in Fig.\ \ref{fig:SQW_InhomoSpectrum}. We observe that a band gap is open around $E=0$, except for $\theta_{\rm max}= 0 \pmod{2\pi}$. A pair of chiral partner zero-energy states appears in the spectral gap whenever $\theta_{\rm min}<0<\theta_{\rm max}$, exponentially localised around the site where $\theta(n)=0$. Similarly, $\pi$ energy states, centred at $\theta(n)=\pi$, appear if $\theta_{\rm min}<\pi<\theta_{\rm max}$.

\begin{figure}[t]
\centering
\includegraphics[width=0.45\textwidth]{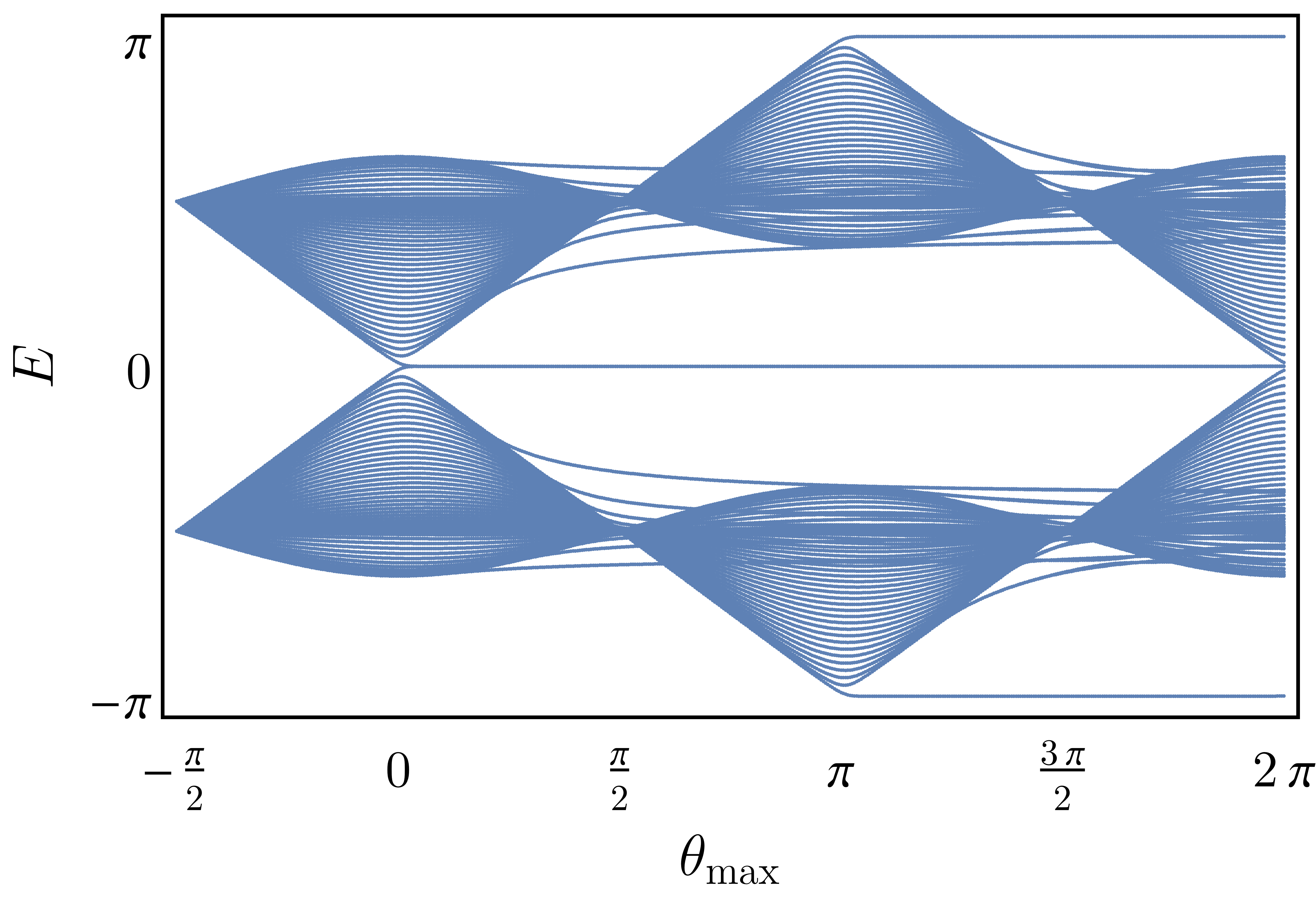}
\caption{Eigenvalues of $\hat H_F'$ for fixed $\theta_{\rm min}=-\pi/2$, as a function of $\theta_{\rm max}$, with open boundary conditions on an $N=300$ site lattice. Pairs of $E\approx 0$ and $E\approx \pi$ states appear in the band gaps and are exponentially localised around the lattice site where $\theta=0$ and $\theta=\pi$, respectively.}
\label{fig:SQW_InhomoSpectrum}
\end{figure}

We will now derive an approximate expression of \^H$_F'$ in the limit $d \ll \xi \ll L$, and use it to explain the origin of the $E\approx 0$ bound states in this model. We will find it useful to go to the continuous limit, where we are able to find an exact eigenstate such that \^H$_F'|\psi\rangle=0$. We will then obtain a discrete ansatz from this solution and evaluate its energy, which we can compare to results from numerical simulations. 

For constant $\theta(n)=\theta$, the band gap closes when $\theta= 0 \pmod{2\pi}$ at $\pm k_{\rm bs}=\pm\pi/(2 d)$, where $d$ is the lattice spacing. For slowly varying $\theta(n)$ near $\theta(n)=0$, we will assume the eigenstates of \^H$_F'$ such that $E\approx 0$ have the form:
\begin{equation}
\label{eq:SQW_FermiState}
|\psi_\pm\rangle=\sum_{n} e^{\pm i k_{\rm bs}  n  d} 
\vec{\varphi}_\pm(n) |n\rangle,
\end{equation}
where $|n\rangle$ is the state which is well localised at site $n$, and $\vec{\varphi}_\pm(n)$ is the envelope function spinor, which we assume to be slowly varying.

When \^H$_S'$ acts on $|\psi_\pm\rangle$, we obtain:
\begin{equation}
\begin{split}
\label{eq:SQW_HPsiDiscrete}
\text{\^H}_S'|\psi_\pm\rangle =&
\pm i J\sum_n  e^{\pm i k_{\rm bs}  n  d}\\
& \sigma_1\cdot\left( \vec{\varphi}_\pm(n+1)-\vec{\varphi}_\pm(n-1) \right)  |n\rangle.
\end{split}
\end{equation}
Assuming that $\vec{\varphi}_\pm(n)$ varies slowly compared to the lattice spacing $d$, we can take the continuous limit of Eq.\ (\ref{eq:SQW_HPsiDiscrete}) by sending $d\rightarrow 0$, $n\rightarrow \infty$ such that $n  d = x$ is constant:
\begin{equation}
\label{eq:SQW_HPsiContinuous}
\text{\^H}_S'|\psi_\pm\rangle\rightarrow
\pm 2 i \tilde{v} \sigma_1 \int dx ~e^{\pm i k_{\rm bs}  x} 
\partial_x  \vec{\varphi}_\pm(x)  |x\rangle,
\end{equation}
where we have introduced the velocity parameter $\tilde{v}=Jd$. The length scale associated with $\partial_x\vec{\varphi}_\pm(x)$ is much larger than the length scale over which $\exp(\pm i k_{\rm bs} x)$ varies, and $\vec{\varphi}_\pm(x)$ decays exponentially away from where $\theta(x)$ vanishes, allowing us to expand \^H$_F'$ to first order in \^H$_S'$ and $\theta(x)$:
\begin{equation}
\label{eq:HToyModel}
\text{\^H}_F'\approx \text{\^H}_{\rm approx}'= \text{\^H}_S' +\text{\^H}_\theta',
\end{equation}
when acting on states near $E=0$ which are localised in the neighbourhood of $\theta(x)=0$. Thus, we have derived a static, approximate Hamiltonian, \^H$_{\rm approx}'$. Using Eq.\ (\ref{eq:SQW_HPsiContinuous}), we can apply this Hamiltonian to $|\psi_\pm\rangle$:
\begin{equation}
\begin{split}
\label{eq:ContinuousHPsi}
\text{\^H}_{\rm approx}'|\psi_\pm\rangle=\int dx ~ e^{\pm i k_{\rm bs} x}\big(& \pm 2 i \tilde{v} \sigma_1\cdot\partial_x \vec{\varphi}(x)\\
& -\theta(x) \sigma_2\cdot \vec{\varphi}(x)\big)|x\rangle.
\end{split}
\end{equation}
It was shown by Jackiw and Rebbi that there always exists a zero solution to the right hand side of Eq.\ (\ref{eq:ContinuousHPsi}) when the sign of $\theta(x)$ is different for $x\rightarrow +\infty$ and $x\rightarrow -\infty$ \cite{Jackiw1976}. When this is the case, \^H$_{\rm approx}'$ has a $E=0$ solution. If $\lim_{x\rightarrow\pm \infty}\text{sign}(\theta(x))=\pm 1$, then $\vec{\varphi}_\pm(x)$ takes the form:
\begin{equation}
\label{eq:JR_envelope}
\vec{\varphi}_\pm(x)=\psi_0 \exp \left(-\frac{1}{2\tilde{v}}\int_0^x dx'\theta(x')\right) |\mp\rangle,
\end{equation}
where we have defined the spin states: $|+\rangle=(1, 0)^T$ and $|-\rangle=(0, 1)^T$, and $\psi_0$ is the normalization of the wave function. In the following, we will restrict our study to $\theta(x)$ varying as:
\begin{equation}
\label{eq:ContinuousBeta}
\theta(x)=\alpha+\beta \tanh\left(\frac{x}{\xi}\right),
\end{equation}
with $\beta>\alpha$ and $\beta>0$. This is the continuous version of Eq.\ \eqref{eq:theta_function}, with $2\alpha=\theta_{\rm max}+\theta_{\rm min}$ and $2\beta=\theta_{\rm max}-\theta_{\rm min}$. From Eqs.\ (\ref{eq:JR_envelope}), (\ref{eq:ContinuousBeta}), we see that $\vec{\varphi}_\pm(x)$ has the form:
\begin{align}
& \vec{\varphi}_\pm(x)=\varphi(x)|\mp\rangle, \\
\label{eq:ContinuousEnvelope}
& \varphi(x)= \psi_0\exp \left(-\frac{\alpha  x}{2\tilde{v}}\right)\cosh\left(\frac{x}{\xi}\right)^{-\beta \xi/(2\tilde{v})}.
\end{align}
By discretising the above result, sending $x\rightarrow n d$, we obtain an ansatz wave function for the two zero energy states that we are observing:
\begin{align}
\label{eq:DiscreteAnsatz}
& |\psi_\pm\rangle=
\sum_n e^{\pm i k_{\rm bs} n d}~ \varphi_n~|n,~ \mp\rangle,\\
\label{eq:DiscreteEnvelope}
& \varphi_n= \psi_0\exp \left(-\frac{\alpha  n}{2 J}\right)\cosh\left(\frac{n  d}{\xi}\right)^{-\beta \xi/(2 J  d)}.
\end{align}

Let's take a moment to recapitulate what we have done so far. We have defined two states, $|\psi_+\rangle$ and $|\psi_-\rangle$, which are centred around the momenta $k=\pm k_{\rm bs}$. We have shown that when $\theta(x)$ changes sign, these states are eigenstates of the Hamiltonian with eigenvalue $E=0$. In the Jackiw-Rebbi model, there is only one such state, which is pinned at $E=0$ by chiral symmetry. In our system, spatially varying $\theta(x)$ leads to terms which mix the states $|\psi_+\rangle$ and $|\psi_-\rangle$, causing them to hybridise and move symmetrically away from $E=0$.

We can find the energy of these hybrid states by studying the eigenvalues of \^H$_{\rm red}'$, the projected Hamiltonian on the basis of states $|\psi_+\rangle$ and $|\psi_-\rangle$.
\begin{equation}
\label{eq:Hreduced}
\text{\^H}_{\rm red}'=\begin{pmatrix}
\langle \psi_+|\text{\^H}_{\rm approx}'|\psi_+\rangle &&
\langle \psi_+|\text{\^H}_{\rm approx}'|\psi_-\rangle \\
\langle \psi_-|\text{\^H}_{\rm approx}'|\psi_+\rangle &&
\langle \psi_-|\text{\^H}_{\rm approx}'|\psi_-\rangle
\end{pmatrix}.
\end{equation}
This matrix has eigenvalues:
\begin{equation}
\label{eq:HredEigenvals}
E_\pm=\pm \sum_n  (-1)^{n} \varphi_n^*
\left(J(\varphi_{n+1}-\varphi_{n-1})+\theta(n) \varphi_n\right).
\end{equation}
The energy $|E_\pm|$ from Eq.\ (\ref{eq:HredEigenvals}) is plotted versus $\xi$ in Fig.\ \ref{fig:JR_energies}. Alongside this estimate of the lowest eigenstate's energy, we have diagonalised \^H$_{\rm approx}'$ and plotted its eigenvalues. Visibly there is a good agreement between the estimate Eq.\ (\ref{eq:HredEigenvals}) and the exact energies, suggesting that our hypothesis was indeed correct, and that the states we are observing are indeed two Jackiw-Rebbi states separated in momentum, the energies of which behave as:
\begin{equation}
\label{eq:Eapprox}
|E_\pm|\approx J \exp\left(-\frac{c \xi}{d}\right),
\end{equation}
where $c$ is a positive constant.

\begin{figure}[t]
\includegraphics[width=0.49\textwidth]{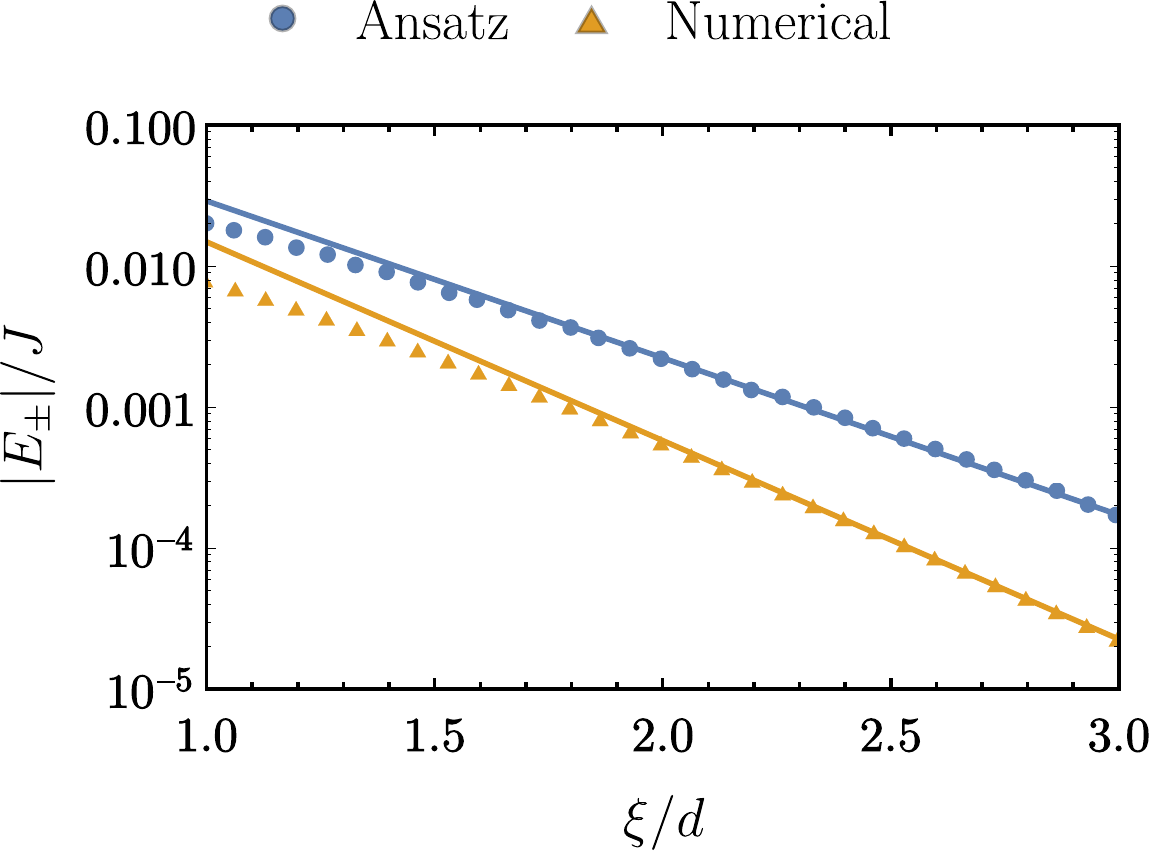}
\caption{Energy of the Jackiw-Rebbi states $|E_\pm|$ versus $\xi$, the inverse rate of change of $\theta(n)$. Blue dots: value of $|E_\pm|$ from Eq.\ (\ref{eq:HredEigenvals}). Yellow triangles: Eigenvalues of \^H$_{\rm approx}'$ closest to $E=0$ (obtained numerically). The lines are exponential fits to the data.
Both these curves approach each other exponentially (which is clear because the scale of the $y$ axis becomes exponentially small). In both cases, $\alpha=-1/4,~\beta=1$.}
\label{fig:JR_energies}
\end{figure}

From the results presented here, we can therefore conclude that the system presents a pair of zero energy bound states when $\theta(n)$ changes sign, despite being in a trivial topological configuration. We find that when $\theta(n)$ changes slowly ($\xi \gg d$), we have two non-overlapping (in $k$ space) bound states, one associated to each Dirac cone. The overlap increases for faster changing $\theta(n)$ (smaller $\xi$), resulting in both states hybridising and moving away from $E=0$ without breaking chiral symmetry. For this reason, these bound states do not benefit from the robustness displayed by topologically bound states.

This type of behaviour is generally observed in systems which present two degenerate energy levels. When the two states are coupled, their energies split symmetrically, proportionally to the matrix element between them (the off diagonal term of Eq.\ (\ref{eq:Hreduced})). In our system, we are able to directly control the tunnelling between these states by modifying the length scale over which the potential varies. Thus our model gives us access to a single parameter which controls the states' degree of hybridisation.

\section{Conclusions}

In Sec.\ \ref{sec:Exp}, we suggested an experiment in which spin $1/2$ cold atoms realise a topologically non-trivial quantum walk. This protocol relies on trapping the atoms in a 1D optical lattice which is spin dependent and has two sites per unit cell. We drive the system by periodically coupling the atoms' spin states. When the dynamics are averaged over one full period of the motion, we see from the operators in Sec.\ \ref{sec:Model} that this driving results in spin-orbit coupling terms. These terms are the key to realising topologically non trivial properties. Specifically, we verified analytically in Sec.\ \ref{sec:TopoProp} that the atomic quantum walk can realise either $E=0$ or $E=\pi$ bound states, which are associated to two separate topological invariants. These states appear when the amplitude of the spin coupling varies spatially; they are bound to the location where one of the topological invariants changes its value.

We showed numerically in Sec.\ \ref{sec:BoundState} that topologically protected bound states can be isolated by performing the atomic quantum walk for a large number of steps ($t/T\approx 50$). Atoms that do not populate the bound state then leave the region of the boundary, such that only an exponentially localised density peak remains at the topological interface. To verify that exponential density peaks such as this one correspond to topological bound states, we suggest exploring the full parameter space, and verifying that exponentially bound states exist whenever the system presents a topological boundary. As an alternative method for identifying topological bound states, we suggested searching for eigenstates of the chiral symmetry operator. These states have a spin distribution which is heavily constrained, allowing us to identify $E=0, \pi$ eigenstates of the Hamiltonian by averaging over spin sensitive measures of the atom's position. By performing the same measure at half time-steps, we saw that we can discriminate between $E=0$ and $E=\pi$ bound states, and thereby measure both of the system's topological invariants.

Finally, we mentioned that it is possible for exponentially bound states to appear at band gap closings where none of the topological invariants change value. In Sec.\ \ref{sec:PairOfJR}, we studied a limit of the atomic quantum walk where this happens; in this case, our model admits two Jackiw-Rebbi states which are separated in quasimomentum. These states can move away from $E=0$ by hybridising, and the degree of their hybridisation is controlled by the rate of change of the spin coupling amplitude. Importantly, these states are not eigenstates of the chiral symmetry operator, meaning that they cannot be confused with topological bound states when performing a spin sensitive measure of the atoms' density distribution.

We have not considered the possibility of spatially varying the tunnelling amplitude $\delta$. This could be done by varying the detuning of spin $\uparrow$ and $\downarrow$ states spatially, as was experimentally realised in Ref.\ \cite{Leder2016a}. By varying $\delta$ spatially, two bound states could be generated at the same location and at the same energy. As can be seen from Fig.\ \ref{fig:phase_diagram}, when the driving is set to $\theta=0$ and $\delta$ changes sign, the winding $\nu_0$ changes by $\Delta\nu_0=2$, implying that there must exist two zero energy bound states at the topological interface. This situation is interesting because, despite the fact that these states overlap spatially, they cannot hybridise.

Aside from this particular parameter regime, it would be interesting to explore the robustness of the topological bound states, when for instance interactions or faults in the periodic driving are introduced. Interactions, as long as they are small and preserve chiral symmetry, should not in principle destroy the system's topological properties. Similarly, the derivation in Appendix \ref{sec:Symmetries} suggests that the bound states should be robust to (perturbative) faults in the periodic driving. It would be interesting to verify theoretically and experimentally to what extent this is the case, as this study would provide a test of how applicable the Floquet description is to real world systems.

\begin{acknowledgments}
The authors would like to thank A.\ Dauphin and G.\ Richardson for fruitful advice and conversations, and L.\ Tarruell for insightful comments. This project was supported by the University of Southampton as host of the Vice-Chancellor Fellowship scheme. We also acknowledge support of the Spanish MINECO (SEVERO OCHOA Grant SEV-2015-0522 and FOQUS FIS2013-46768), the Generalitat de Catalunya (SGR 874), Fundaci\'o Privada Cellex, and ERC AdG OSYRIS. P.\ M.\ acknowledges support from the Ram\'on y Cajal programme.
\end{acknowledgments}

\bibliographystyle{apsrev4-1_our_style}
\bibliography{library,Algorithms_citations}

\begin{thebibliography}{43}%
\makeatletter
\providecommand \@ifxundefined [1]{%
 \@ifx{#1\undefined}
}%
\providecommand \@ifnum [1]{%
 \ifnum #1\expandafter \@firstoftwo
 \else \expandafter \@secondoftwo
 \fi
}%
\providecommand \@ifx [1]{%
 \ifx #1\expandafter \@firstoftwo
 \else \expandafter \@secondoftwo
 \fi
}%
\providecommand \natexlab [1]{#1}%
\providecommand \enquote  [1]{``#1''}%
\providecommand \bibnamefont  [1]{#1}%
\providecommand \bibfnamefont [1]{#1}%
\providecommand \citenamefont [1]{#1}%
\providecommand \href@noop [0]{\@secondoftwo}%
\providecommand \href [0]{\begingroup \@sanitize@url \@href}%
\providecommand \@href[1]{\@@startlink{#1}\@@href}%
\providecommand \@@href[1]{\endgroup#1\@@endlink}%
\providecommand \@sanitize@url [0]{\catcode `\\12\catcode `\$12\catcode
  `\&12\catcode `\#12\catcode `\^12\catcode `\_12\catcode `\%12\relax}%
\providecommand \@@startlink[1]{}%
\providecommand \@@endlink[0]{}%
\providecommand \url  [0]{\begingroup\@sanitize@url \@url }%
\providecommand \@url [1]{\endgroup\@href {#1}{\urlprefix }}%
\providecommand \urlprefix  [0]{URL }%
\providecommand \Eprint [0]{\href }%
\providecommand \doibase [0]{http://dx.doi.org/}%
\providecommand \selectlanguage [0]{\@gobble}%
\providecommand \bibinfo  [0]{\@secondoftwo}%
\providecommand \bibfield  [0]{\@secondoftwo}%
\providecommand \translation [1]{[#1]}%
\providecommand \BibitemOpen [0]{}%
\providecommand \bibitemStop [0]{}%
\providecommand \bibitemNoStop [0]{.\EOS\space}%
\providecommand \EOS [0]{\spacefactor3000\relax}%
\providecommand \BibitemShut  [1]{\csname bibitem#1\endcsname}%
\let\auto@bib@innerbib\@empty
\bibitem [{\citenamefont {Chisaki}\ \emph {et~al.}(2010)\citenamefont
  {Chisaki}, \citenamefont {Konno},\ and\ \citenamefont
  {Segawa}}]{Chisaki2012}%
  \BibitemOpen
  \bibfield  {author} {\bibinfo {author} {\bibfnamefont {K.}~\bibnamefont
  {Chisaki}}, \bibinfo {author} {\bibfnamefont {N.}~\bibnamefont {Konno}}, \
  and\ \bibinfo {author} {\bibfnamefont {E.}~\bibnamefont {Segawa}},\
  }\bibfield  {title} {\bibinfo {title} {\emph {{Limit Theorems for the
  Discrete-Time Quantum Walk on a Graph with Joined Half Lines}}},\ }\href
  {http://arxiv.org/abs/1009.1306} {\bibfield  {journal} {\bibinfo  {journal}
  {Quantum Information and Computation}\ }\textbf {\bibinfo {volume} {12}},\
  \bibinfo {pages} {0314}\  (\bibinfo {year} {2010})}\BibitemShut {NoStop}%
\bibitem [{\citenamefont {Childs}\ \emph {et~al.}(2003)\citenamefont {Childs},
  \citenamefont {Cleve}, \citenamefont {Deotto}, \citenamefont {Farhi},
  \citenamefont {Gutmann},\ and\ \citenamefont {Spielman}}]{Childs2003}%
  \BibitemOpen
  \bibfield  {author} {\bibinfo {author} {\bibfnamefont {A.~M.}\ \bibnamefont
  {Childs}}, \bibinfo {author} {\bibfnamefont {R.}~\bibnamefont {Cleve}},
  \bibinfo {author} {\bibfnamefont {E.}~\bibnamefont {Deotto}}, \bibinfo
  {author} {\bibfnamefont {E.}~\bibnamefont {Farhi}}, \bibinfo {author}
  {\bibfnamefont {S.}~\bibnamefont {Gutmann}}, \ and\ \bibinfo {author}
  {\bibfnamefont {D.~A.}\ \bibnamefont {Spielman}},\ }\bibfield  {title}
  {\bibinfo {title} {\emph {Exponential Algorithmic Speedup by a Quantum
  Walk}},\ }in\ \href {\doibase 10.1145/780542.780552} {\emph {\bibinfo
  {booktitle} {Proceedings of the Thirty-fifth Annual ACM Symposium on Theory
  of Computing}}},\ \bibinfo {series and number} {STOC '03}\ (\bibinfo
  {publisher} {ACM},\ \bibinfo {address} {New York, NY, USA},\ \bibinfo {year}
  {2003})\ \ pp.\ \bibinfo {pages} {59--68}\BibitemShut {NoStop}%
\bibitem [{\citenamefont {Ambainis}(2004)}]{Ambainis04quantumwalk}%
  \BibitemOpen
  \bibfield  {author} {\bibinfo {author} {\bibfnamefont {A.}~\bibnamefont
  {Ambainis}},\ }\bibfield  {title} {\bibinfo {title} {\emph {Quantum walk
  algorithms for element distinctness}},\ }in\ \href {\doibase
  10.1109/FOCS.2004.54} {\emph {\bibinfo {booktitle} {45th Annual IEEE
  Symposium on Foundations of Computer Science, OCT 17-19, 2004. IEEE Computer
  Society Press, Los Alamitos, CA}}}\ (\bibinfo {year} {2004})\ \ pp.\ \bibinfo
  {pages} {22--31}\BibitemShut {NoStop}%
\bibitem [{\citenamefont {Magniez}\ \emph {et~al.}(2005)\citenamefont
  {Magniez}, \citenamefont {Santha},\ and\ \citenamefont
  {Szegedy}}]{Magniez2005}%
  \BibitemOpen
  \bibfield  {author} {\bibinfo {author} {\bibfnamefont {F.}~\bibnamefont
  {Magniez}}, \bibinfo {author} {\bibfnamefont {M.}~\bibnamefont {Santha}}, \
  and\ \bibinfo {author} {\bibfnamefont {M.}~\bibnamefont {Szegedy}},\
  }\bibfield  {title} {\bibinfo {title} {\emph {Quantum Algorithms for the
  Triangle Problem}},\ }in\ \href {\doibase 10.1137/050643684} {\emph {\bibinfo
  {booktitle} {Proceedings of the Sixteenth Annual ACM-SIAM Symposium on
  Discrete Algorithms}}},\ \bibinfo {series and number} {SODA '05}\ (\bibinfo
  {publisher} {Society for Industrial and Applied Mathematics},\ \bibinfo
  {address} {Philadelphia, PA, USA},\ \bibinfo {year} {2005})\ \ pp.\ \bibinfo
  {pages} {1109--1117}\BibitemShut {NoStop}%
\bibitem [{\citenamefont {Farhi}\ \emph {et~al.}(2008)\citenamefont {Farhi},
  \citenamefont {Goldstone},\ and\ \citenamefont {Gutmann}}]{Farhi07aquantum}%
  \BibitemOpen
  \bibfield  {author} {\bibinfo {author} {\bibfnamefont {E.}~\bibnamefont
  {Farhi}}, \bibinfo {author} {\bibfnamefont {J.}~\bibnamefont {Goldstone}}, \
  and\ \bibinfo {author} {\bibfnamefont {S.}~\bibnamefont {Gutmann}},\
  }\bibfield  {title} {\bibinfo {title} {\emph {A Quantum Algorithm for the
  Hamiltonian NAND Tree}},\ }\href {\doibase 10.4086/toc.2008.v004a008}
  {\bibfield  {journal} {\bibinfo  {journal} {Theory of Computing}\ }\textbf
  {\bibinfo {volume} {4}},\ \bibinfo {pages} {169}\  (\bibinfo {year}
  {2008})}\BibitemShut {NoStop}%
\bibitem [{\citenamefont {Feynman}\ and\ \citenamefont
  {Hibbs}(2012)}]{Feynman2012}%
  \BibitemOpen
  \bibfield  {author} {\bibinfo {author} {\bibfnamefont {R.}~\bibnamefont
  {Feynman}}\ and\ \bibinfo {author} {\bibfnamefont {A.}~\bibnamefont
  {Hibbs}},\ }\href {http://store.doverpublications.com/0486477223.html} {\emph
  {\bibinfo {title} {Quantum Mechanics and Path Integrals: Emended Edition}}}\
  (\bibinfo  {publisher} {Dover Publications, Incorporated},\ \bibinfo {year}
  {2012})\BibitemShut {NoStop}%
\bibitem [{\citenamefont {Strauch}(2006)}]{Strauch2006}%
  \BibitemOpen
  \bibfield  {author} {\bibinfo {author} {\bibfnamefont {F.~W.}\ \bibnamefont
  {Strauch}},\ }\bibfield  {title} {\bibinfo {title} {\emph {{Relativistic
  quantum walks}}},\ }\href {\doibase 10.1103/PhysRevA.73.054302} {\bibfield
  {journal} {\bibinfo  {journal} {Physical Review A}\ }\textbf {\bibinfo
  {volume} {73}},\ \bibinfo {pages} {054302}\  (\bibinfo {year}
  {2006})}\BibitemShut {NoStop}%
\bibitem [{\citenamefont {Oka}\ \emph {et~al.}(2004)\citenamefont {Oka},
  \citenamefont {Konno}, \citenamefont {Arita},\ and\ \citenamefont
  {Aoki}}]{Oka2005}%
  \BibitemOpen
  \bibfield  {author} {\bibinfo {author} {\bibfnamefont {T.}~\bibnamefont
  {Oka}}, \bibinfo {author} {\bibfnamefont {N.}~\bibnamefont {Konno}}, \bibinfo
  {author} {\bibfnamefont {R.}~\bibnamefont {Arita}}, \ and\ \bibinfo {author}
  {\bibfnamefont {H.}~\bibnamefont {Aoki}},\ }\bibfield  {title} {\bibinfo
  {title} {\emph {{Breakdown of an Electric-Field Driven System: a Mapping to a
  Quantum Walk}}},\ }\href {\doibase 10.1103/PhysRevLett.94.100602} {\bibfield
  {journal} {\bibinfo  {journal} {Physical Review Letters}\ }\textbf {\bibinfo
  {volume} {94}},\ \bibinfo {pages} {100602}\  (\bibinfo {year}
  {2004})}\BibitemShut {NoStop}%
\bibitem [{\citenamefont {Peruzzo}\ \emph {et~al.}(2010)\citenamefont
  {Peruzzo}, \citenamefont {Lobino}, \citenamefont {Matthews}, \citenamefont
  {Matsuda}, \citenamefont {Politi}, \citenamefont {Poulios}, \citenamefont
  {Zhou}, \citenamefont {Lahini}, \citenamefont {Ismail}, \citenamefont
  {Worhoff}, \citenamefont {Bromberg}, \citenamefont {Silberberg},
  \citenamefont {Thompson},\ and\ \citenamefont {OBrien}}]{Peruzzo2010}%
  \BibitemOpen
  \bibfield  {author} {\bibinfo {author} {\bibfnamefont {A.}~\bibnamefont
  {Peruzzo}}, \bibinfo {author} {\bibfnamefont {M.}~\bibnamefont {Lobino}},
  \bibinfo {author} {\bibfnamefont {J.~C.~F.}\ \bibnamefont {Matthews}},
  \bibinfo {author} {\bibfnamefont {N.}~\bibnamefont {Matsuda}}, \bibinfo
  {author} {\bibfnamefont {A.}~\bibnamefont {Politi}}, \bibinfo {author}
  {\bibfnamefont {K.}~\bibnamefont {Poulios}}, \bibinfo {author} {\bibfnamefont
  {X.-Q.}\ \bibnamefont {Zhou}}, \bibinfo {author} {\bibfnamefont
  {Y.}~\bibnamefont {Lahini}}, \bibinfo {author} {\bibfnamefont
  {N.}~\bibnamefont {Ismail}}, \bibinfo {author} {\bibfnamefont
  {K.}~\bibnamefont {Worhoff}}, \bibinfo {author} {\bibfnamefont
  {Y.}~\bibnamefont {Bromberg}}, \bibinfo {author} {\bibfnamefont
  {Y.}~\bibnamefont {Silberberg}}, \bibinfo {author} {\bibfnamefont {M.~G.}\
  \bibnamefont {Thompson}}, \ and\ \bibinfo {author} {\bibfnamefont {J.~L.}\
  \bibnamefont {OBrien}},\ }\bibfield  {title} {\bibinfo {title} {\emph
  {{Quantum Walks of Correlated Photons}}},\ }\href {\doibase
  10.1126/science.1193515} {\bibfield  {journal} {\bibinfo  {journal}
  {Science}\ }\textbf {\bibinfo {volume} {329}},\ \bibinfo {pages} {1500}\
  (\bibinfo {year} {2010})}\BibitemShut {NoStop}%
\bibitem [{\citenamefont {Schreiber}\ \emph {et~al.}(2011)\citenamefont
  {Schreiber}, \citenamefont {Cassemiro}, \citenamefont {Poto{\v{c}}ek},
  \citenamefont {G{\'{a}}bris}, \citenamefont {Jex},\ and\ \citenamefont
  {Silberhorn}}]{Schreiber2011}%
  \BibitemOpen
  \bibfield  {author} {\bibinfo {author} {\bibfnamefont {A.}~\bibnamefont
  {Schreiber}}, \bibinfo {author} {\bibfnamefont {K.~N.}\ \bibnamefont
  {Cassemiro}}, \bibinfo {author} {\bibfnamefont {V.}~\bibnamefont
  {Poto{\v{c}}ek}}, \bibinfo {author} {\bibfnamefont {A.}~\bibnamefont
  {G{\'{a}}bris}}, \bibinfo {author} {\bibfnamefont {I.}~\bibnamefont {Jex}}, \
  and\ \bibinfo {author} {\bibfnamefont {C.}~\bibnamefont {Silberhorn}},\
  }\bibfield  {title} {\bibinfo {title} {\emph {{Decoherence and Disorder in
  Quantum Walks: From Ballistic Spread to Localization}}},\ }\href {\doibase
  10.1103/PhysRevLett.106.180403} {\bibfield  {journal} {\bibinfo  {journal}
  {Physical Review Letters}\ }\textbf {\bibinfo {volume} {106}},\ \bibinfo
  {pages} {180403}\  (\bibinfo {year} {2011})}\BibitemShut {NoStop}%
\bibitem [{\citenamefont {Schreiber}\ \emph {et~al.}(2012)\citenamefont
  {Schreiber}, \citenamefont {Gabris}, \citenamefont {Rohde}, \citenamefont
  {Laiho}, \citenamefont {Stefanak}, \citenamefont {Potocek}, \citenamefont
  {Hamilton}, \citenamefont {Jex},\ and\ \citenamefont
  {Silberhorn}}]{Schreiber2012}%
  \BibitemOpen
  \bibfield  {author} {\bibinfo {author} {\bibfnamefont {A.}~\bibnamefont
  {Schreiber}}, \bibinfo {author} {\bibfnamefont {A.}~\bibnamefont {Gabris}},
  \bibinfo {author} {\bibfnamefont {P.~P.}\ \bibnamefont {Rohde}}, \bibinfo
  {author} {\bibfnamefont {K.}~\bibnamefont {Laiho}}, \bibinfo {author}
  {\bibfnamefont {M.}~\bibnamefont {Stefanak}}, \bibinfo {author}
  {\bibfnamefont {V.}~\bibnamefont {Potocek}}, \bibinfo {author} {\bibfnamefont
  {C.}~\bibnamefont {Hamilton}}, \bibinfo {author} {\bibfnamefont
  {I.}~\bibnamefont {Jex}}, \ and\ \bibinfo {author} {\bibfnamefont
  {C.}~\bibnamefont {Silberhorn}},\ }\bibfield  {title} {\bibinfo {title}
  {\emph {{A 2D Quantum Walk Simulation of Two-Particle Dynamics}}},\ }\href
  {\doibase 10.1126/science.1218448} {\bibfield  {journal} {\bibinfo  {journal}
  {Science}\ }\textbf {\bibinfo {volume} {336}},\ \bibinfo {pages} {55}\
  (\bibinfo {year} {2012})}\BibitemShut {NoStop}%
\bibitem [{\citenamefont {Kitagawa}\ \emph {et~al.}(2012)\citenamefont
  {Kitagawa}, \citenamefont {Broome}, \citenamefont {Fedrizzi}, \citenamefont
  {Rudner}, \citenamefont {Berg}, \citenamefont {Kassal}, \citenamefont
  {Aspuru-Guzik}, \citenamefont {Demler},\ and\ \citenamefont
  {White}}]{Kitagawa2012a}%
  \BibitemOpen
  \bibfield  {author} {\bibinfo {author} {\bibfnamefont {T.}~\bibnamefont
  {Kitagawa}}, \bibinfo {author} {\bibfnamefont {M.~A.}\ \bibnamefont
  {Broome}}, \bibinfo {author} {\bibfnamefont {A.}~\bibnamefont {Fedrizzi}},
  \bibinfo {author} {\bibfnamefont {M.~S.}\ \bibnamefont {Rudner}}, \bibinfo
  {author} {\bibfnamefont {E.}~\bibnamefont {Berg}}, \bibinfo {author}
  {\bibfnamefont {I.}~\bibnamefont {Kassal}}, \bibinfo {author} {\bibfnamefont
  {A.}~\bibnamefont {Aspuru-Guzik}}, \bibinfo {author} {\bibfnamefont
  {E.}~\bibnamefont {Demler}}, \ and\ \bibinfo {author} {\bibfnamefont {A.~G.}\
  \bibnamefont {White}},\ }\bibfield  {title} {\bibinfo {title} {\emph
  {{Observation of topologically protected bound states in photonic quantum
  walks}}},\ }\href {\doibase 10.1038/ncomms1872} {\bibfield  {journal}
  {\bibinfo  {journal} {Nature Communications}\ }\textbf {\bibinfo {volume}
  {3}},\ \bibinfo {pages} {882}\  (\bibinfo {year} {2012})}\BibitemShut
  {NoStop}%
\bibitem [{\citenamefont {Crespi}\ \emph {et~al.}(2013)\citenamefont {Crespi},
  \citenamefont {Osellame}, \citenamefont {Ramponi}, \citenamefont
  {Giovannetti}, \citenamefont {Fazio}, \citenamefont {Sansoni}, \citenamefont
  {{De Nicola}}, \citenamefont {Sciarrino},\ and\ \citenamefont
  {Mataloni}}]{Crespi2013}%
  \BibitemOpen
  \bibfield  {author} {\bibinfo {author} {\bibfnamefont {A.}~\bibnamefont
  {Crespi}}, \bibinfo {author} {\bibfnamefont {R.}~\bibnamefont {Osellame}},
  \bibinfo {author} {\bibfnamefont {R.}~\bibnamefont {Ramponi}}, \bibinfo
  {author} {\bibfnamefont {V.}~\bibnamefont {Giovannetti}}, \bibinfo {author}
  {\bibfnamefont {R.}~\bibnamefont {Fazio}}, \bibinfo {author} {\bibfnamefont
  {L.}~\bibnamefont {Sansoni}}, \bibinfo {author} {\bibfnamefont
  {F.}~\bibnamefont {{De Nicola}}}, \bibinfo {author} {\bibfnamefont
  {F.}~\bibnamefont {Sciarrino}}, \ and\ \bibinfo {author} {\bibfnamefont
  {P.}~\bibnamefont {Mataloni}},\ }\bibfield  {title} {\bibinfo {title} {\emph
  {{Anderson localization of entangled photons in an integrated quantum
  walk}}},\ }\href {\doibase 10.1038/nphoton.2013.26} {\bibfield  {journal}
  {\bibinfo  {journal} {Nature Photonics}\ }\textbf {\bibinfo {volume} {7}},\
  \bibinfo {pages} {322}\  (\bibinfo {year} {2013})}\BibitemShut {NoStop}%
\bibitem [{\citenamefont {Rechtsman}\ \emph {et~al.}(2013)\citenamefont
  {Rechtsman}, \citenamefont {Plotnik}, \citenamefont {Zeuner}, \citenamefont
  {Song}, \citenamefont {Chen}, \citenamefont {Szameit},\ and\ \citenamefont
  {Segev}}]{Rechtsman2013}%
  \BibitemOpen
  \bibfield  {author} {\bibinfo {author} {\bibfnamefont {M.~C.}\ \bibnamefont
  {Rechtsman}}, \bibinfo {author} {\bibfnamefont {Y.}~\bibnamefont {Plotnik}},
  \bibinfo {author} {\bibfnamefont {J.~M.}\ \bibnamefont {Zeuner}}, \bibinfo
  {author} {\bibfnamefont {D.}~\bibnamefont {Song}}, \bibinfo {author}
  {\bibfnamefont {Z.}~\bibnamefont {Chen}}, \bibinfo {author} {\bibfnamefont
  {A.}~\bibnamefont {Szameit}}, \ and\ \bibinfo {author} {\bibfnamefont
  {M.}~\bibnamefont {Segev}},\ }\bibfield  {title} {\bibinfo {title} {\emph
  {{Topological creation and destruction of edge states in photonic
  graphene}}},\ }\href {\doibase 10.1103/PhysRevLett.111.103901} {\bibfield
  {journal} {\bibinfo  {journal} {Physical Review Letters}\ }\textbf {\bibinfo
  {volume} {111}},\ \bibinfo {pages} {103901}\  (\bibinfo {year}
  {2013})}\BibitemShut {NoStop}%
\bibitem [{\citenamefont {Cardano}\ \emph
  {et~al.}(2015{\natexlab{a}})\citenamefont {Cardano}, \citenamefont {Massa},
  \citenamefont {Qassim}, \citenamefont {Karimi}, \citenamefont {Slussarenko},
  \citenamefont {Paparo}, \citenamefont {de~Lisio}, \citenamefont {Sciarrino},
  \citenamefont {Santamato}, \citenamefont {Boyd},\ and\ \citenamefont
  {Marrucci}}]{Cardano2015}%
  \BibitemOpen
  \bibfield  {author} {\bibinfo {author} {\bibfnamefont {F.}~\bibnamefont
  {Cardano}}, \bibinfo {author} {\bibfnamefont {F.}~\bibnamefont {Massa}},
  \bibinfo {author} {\bibfnamefont {H.}~\bibnamefont {Qassim}}, \bibinfo
  {author} {\bibfnamefont {E.}~\bibnamefont {Karimi}}, \bibinfo {author}
  {\bibfnamefont {S.}~\bibnamefont {Slussarenko}}, \bibinfo {author}
  {\bibfnamefont {D.}~\bibnamefont {Paparo}}, \bibinfo {author} {\bibfnamefont
  {C.}~\bibnamefont {de~Lisio}}, \bibinfo {author} {\bibfnamefont
  {F.}~\bibnamefont {Sciarrino}}, \bibinfo {author} {\bibfnamefont
  {E.}~\bibnamefont {Santamato}}, \bibinfo {author} {\bibfnamefont {R.~W.}\
  \bibnamefont {Boyd}}, \ and\ \bibinfo {author} {\bibfnamefont
  {L.}~\bibnamefont {Marrucci}},\ }\bibfield  {title} {\bibinfo {title} {\emph
  {{Quantum walks and wavepacket dynamics on a lattice with twisted
  photons}}},\ }\href {\doibase 10.1126/sciadv.1500087} {\bibfield  {journal}
  {\bibinfo  {journal} {Science Advances}\ }\textbf {\bibinfo {volume} {1}},\
  \bibinfo {pages} {e1500087}\  (\bibinfo {year}
  {2015}{\natexlab{a}})}\BibitemShut {NoStop}%
\bibitem [{\citenamefont {Cardano}\ \emph
  {et~al.}(2015{\natexlab{b}})\citenamefont {Cardano}, \citenamefont {Maffei},
  \citenamefont {Massa}, \citenamefont {Piccirillo}, \citenamefont {de~Lisio},
  \citenamefont {{De Filippis}}, \citenamefont {Cataudella}, \citenamefont
  {Santamato},\ and\ \citenamefont {Marrucci}}]{Cardano2015a}%
  \BibitemOpen
  \bibfield  {author} {\bibinfo {author} {\bibfnamefont {F.}~\bibnamefont
  {Cardano}}, \bibinfo {author} {\bibfnamefont {M.}~\bibnamefont {Maffei}},
  \bibinfo {author} {\bibfnamefont {F.}~\bibnamefont {Massa}}, \bibinfo
  {author} {\bibfnamefont {B.}~\bibnamefont {Piccirillo}}, \bibinfo {author}
  {\bibfnamefont {C.}~\bibnamefont {de~Lisio}}, \bibinfo {author}
  {\bibfnamefont {G.}~\bibnamefont {{De Filippis}}}, \bibinfo {author}
  {\bibfnamefont {V.}~\bibnamefont {Cataudella}}, \bibinfo {author}
  {\bibfnamefont {E.}~\bibnamefont {Santamato}}, \ and\ \bibinfo {author}
  {\bibfnamefont {L.}~\bibnamefont {Marrucci}},\ }\bibfield  {title} {\bibinfo
  {title} {\emph {{Dynamical moments reveal a topological quantum transition in
  a photonic quantum walk}}},\ }\href {http://arxiv.org/abs/1507.01785}
  {\bibfield  {journal} {\bibinfo  {journal} {arXiv:1507.01785}\ }\  (\bibinfo
  {year} {2015}{\natexlab{b}})}\BibitemShut {NoStop}%
\bibitem [{\citenamefont {Karski}\ \emph {et~al.}(2009)\citenamefont {Karski},
  \citenamefont {Forster}, \citenamefont {Choi}, \citenamefont {Steffen},
  \citenamefont {Alt}, \citenamefont {Meschede},\ and\ \citenamefont
  {Widera}}]{Karski2009}%
  \BibitemOpen
  \bibfield  {author} {\bibinfo {author} {\bibfnamefont {M.}~\bibnamefont
  {Karski}}, \bibinfo {author} {\bibfnamefont {L.}~\bibnamefont {Forster}},
  \bibinfo {author} {\bibfnamefont {J.-M.}\ \bibnamefont {Choi}}, \bibinfo
  {author} {\bibfnamefont {A.}~\bibnamefont {Steffen}}, \bibinfo {author}
  {\bibfnamefont {W.}~\bibnamefont {Alt}}, \bibinfo {author} {\bibfnamefont
  {D.}~\bibnamefont {Meschede}}, \ and\ \bibinfo {author} {\bibfnamefont
  {A.}~\bibnamefont {Widera}},\ }\bibfield  {title} {\bibinfo {title} {\emph
  {{Quantum Walk in Position Space with Single Optically Trapped Atoms}}},\
  }\href {\doibase 10.1126/science.1174436} {\bibfield  {journal} {\bibinfo
  {journal} {Science}\ }\textbf {\bibinfo {volume} {325}},\ \bibinfo {pages}
  {174}\  (\bibinfo {year} {2009})}\BibitemShut {NoStop}%
\bibitem [{\citenamefont {Genske}\ \emph {et~al.}(2013)\citenamefont {Genske},
  \citenamefont {Alt}, \citenamefont {Steffen}, \citenamefont {Werner},
  \citenamefont {Werner}, \citenamefont {Meschede},\ and\ \citenamefont
  {Alberti}}]{Genske2013}%
  \BibitemOpen
  \bibfield  {author} {\bibinfo {author} {\bibfnamefont {M.}~\bibnamefont
  {Genske}}, \bibinfo {author} {\bibfnamefont {W.}~\bibnamefont {Alt}},
  \bibinfo {author} {\bibfnamefont {A.}~\bibnamefont {Steffen}}, \bibinfo
  {author} {\bibfnamefont {A.~H.}\ \bibnamefont {Werner}}, \bibinfo {author}
  {\bibfnamefont {R.~F.}\ \bibnamefont {Werner}}, \bibinfo {author}
  {\bibfnamefont {D.}~\bibnamefont {Meschede}}, \ and\ \bibinfo {author}
  {\bibfnamefont {A.}~\bibnamefont {Alberti}},\ }\bibfield  {title} {\bibinfo
  {title} {\emph {{Electric quantum walks with individual atoms}}},\ }\href
  {\doibase 10.1103/PhysRevLett.110.190601} {\bibfield  {journal} {\bibinfo
  {journal} {Physical Review Letters}\ }\textbf {\bibinfo {volume} {110}},\
  \bibinfo {pages} {190601}\  (\bibinfo {year} {2013})}\BibitemShut {NoStop}%
\bibitem [{\citenamefont {Preiss}\ \emph {et~al.}(2015)\citenamefont {Preiss},
  \citenamefont {Tai}, \citenamefont {Lukin}, \citenamefont {Rispoli},
  \citenamefont {Zupancic}, \citenamefont {Lahini}, \citenamefont {Islam},\
  and\ \citenamefont {Greiner}}]{Preiss2014}%
  \BibitemOpen
  \bibfield  {author} {\bibinfo {author} {\bibfnamefont {P.~M.}\ \bibnamefont
  {Preiss}}, \bibinfo {author} {\bibfnamefont {M.~E.}\ \bibnamefont {Tai}},
  \bibinfo {author} {\bibfnamefont {A.}~\bibnamefont {Lukin}}, \bibinfo
  {author} {\bibfnamefont {M.}~\bibnamefont {Rispoli}}, \bibinfo {author}
  {\bibfnamefont {P.}~\bibnamefont {Zupancic}}, \bibinfo {author}
  {\bibfnamefont {Y.}~\bibnamefont {Lahini}}, \bibinfo {author} {\bibfnamefont
  {R.}~\bibnamefont {Islam}}, \ and\ \bibinfo {author} {\bibfnamefont
  {M.}~\bibnamefont {Greiner}},\ }\bibfield  {title} {\bibinfo {title} {\emph
  {{Strongly correlated quantum walks in optical lattices}}},\ }\href {\doibase
  10.1126/science.1260364} {\bibfield  {journal} {\bibinfo  {journal}
  {Science}\ }\textbf {\bibinfo {volume} {347}},\ \bibinfo {pages} {1229}\
  (\bibinfo {year} {2015})}\BibitemShut {NoStop}%
\bibitem [{\citenamefont {Robens}\ \emph
  {et~al.}(2015{\natexlab{a}})\citenamefont {Robens}, \citenamefont {Brakhane},
  \citenamefont {Meschede},\ and\ \citenamefont {Alberti}}]{Robens2015}%
  \BibitemOpen
  \bibfield  {author} {\bibinfo {author} {\bibfnamefont {C.}~\bibnamefont
  {Robens}}, \bibinfo {author} {\bibfnamefont {S.}~\bibnamefont {Brakhane}},
  \bibinfo {author} {\bibfnamefont {D.}~\bibnamefont {Meschede}}, \ and\
  \bibinfo {author} {\bibfnamefont {A.}~\bibnamefont {Alberti}},\ }\bibfield
  {title} {\bibinfo {title} {\emph {{Quantum Walks With Neutral Atoms: Quantum
  Interference Effects of One and Two Particles}}},\ }\href
  {http://arxiv.org/abs/1511.03569} {\bibfield  {journal} {\bibinfo  {journal}
  {Proceedings of the XXII International Conference ICOLS}\ }\  (\bibinfo
  {year} {2015}{\natexlab{a}})}\BibitemShut {NoStop}%
\bibitem [{\citenamefont {Robens}\ \emph
  {et~al.}(2015{\natexlab{b}})\citenamefont {Robens}, \citenamefont {Alt},
  \citenamefont {Meschede}, \citenamefont {Emary},\ and\ \citenamefont
  {Alberti}}]{Robens2015a}%
  \BibitemOpen
  \bibfield  {author} {\bibinfo {author} {\bibfnamefont {C.}~\bibnamefont
  {Robens}}, \bibinfo {author} {\bibfnamefont {W.}~\bibnamefont {Alt}},
  \bibinfo {author} {\bibfnamefont {D.}~\bibnamefont {Meschede}}, \bibinfo
  {author} {\bibfnamefont {C.}~\bibnamefont {Emary}}, \ and\ \bibinfo {author}
  {\bibfnamefont {A.}~\bibnamefont {Alberti}},\ }\bibfield  {title} {\bibinfo
  {title} {\emph {{Ideal Negative Measurements in Quantum Walks Disprove
  Theories Based on Classical Trajectories}}},\ }\href {\doibase
  10.1103/PhysRevX.5.011003} {\bibfield  {journal} {\bibinfo  {journal}
  {Physical Review X}\ }\textbf {\bibinfo {volume} {5}},\ \bibinfo {pages}
  {011003}\  (\bibinfo {year} {2015}{\natexlab{b}})}\BibitemShut {NoStop}%
\bibitem [{\citenamefont {Meier}\ \emph {et~al.}(2016)\citenamefont {Meier},
  \citenamefont {An},\ and\ \citenamefont {Gadway}}]{Meier2016}%
  \BibitemOpen
  \bibfield  {author} {\bibinfo {author} {\bibfnamefont {E.~J.}\ \bibnamefont
  {Meier}}, \bibinfo {author} {\bibfnamefont {F.~A.}\ \bibnamefont {An}}, \
  and\ \bibinfo {author} {\bibfnamefont {B.}~\bibnamefont {Gadway}},\
  }\bibfield  {title} {\bibinfo {title} {\emph {{Atom optics simulator of
  lattice transport phenomena}}},\ }\href {http://arxiv.org/abs/1601.05785}
  {\bibfield  {journal} {\bibinfo  {journal} {arXiv:1601.05785}\ }\  (\bibinfo
  {year} {2016})}\BibitemShut {NoStop}%
\bibitem [{\citenamefont {Schmitz}\ \emph {et~al.}(2009)\citenamefont
  {Schmitz}, \citenamefont {Matjeschk}, \citenamefont {Schneider},
  \citenamefont {Glueckert}, \citenamefont {Enderlein}, \citenamefont {Huber},\
  and\ \citenamefont {Schaetz}}]{Schmitz2009}%
  \BibitemOpen
  \bibfield  {author} {\bibinfo {author} {\bibfnamefont {H.}~\bibnamefont
  {Schmitz}}, \bibinfo {author} {\bibfnamefont {R.}~\bibnamefont {Matjeschk}},
  \bibinfo {author} {\bibfnamefont {C.}~\bibnamefont {Schneider}}, \bibinfo
  {author} {\bibfnamefont {J.}~\bibnamefont {Glueckert}}, \bibinfo {author}
  {\bibfnamefont {M.}~\bibnamefont {Enderlein}}, \bibinfo {author}
  {\bibfnamefont {T.}~\bibnamefont {Huber}}, \ and\ \bibinfo {author}
  {\bibfnamefont {T.}~\bibnamefont {Schaetz}},\ }\bibfield  {title} {\bibinfo
  {title} {\emph {{Quantum walk of a trapped ion in phase space}}},\ }\href
  {\doibase 10.1103/PhysRevLett.103.090504} {\bibfield  {journal} {\bibinfo
  {journal} {Physical Review Letters}\ }\textbf {\bibinfo {volume} {103}},\
  \bibinfo {pages} {090504}\  (\bibinfo {year} {2009})}\BibitemShut {NoStop}%
\bibitem [{\citenamefont {Z{\"{a}}hringer}\ \emph {et~al.}(2010)\citenamefont
  {Z{\"{a}}hringer}, \citenamefont {Kirchmair}, \citenamefont {Gerritsma},
  \citenamefont {Solano}, \citenamefont {Blatt},\ and\ \citenamefont
  {Roos}}]{Zahringer2010}%
  \BibitemOpen
  \bibfield  {author} {\bibinfo {author} {\bibfnamefont {F.}~\bibnamefont
  {Z{\"{a}}hringer}}, \bibinfo {author} {\bibfnamefont {G.}~\bibnamefont
  {Kirchmair}}, \bibinfo {author} {\bibfnamefont {R.}~\bibnamefont
  {Gerritsma}}, \bibinfo {author} {\bibfnamefont {E.}~\bibnamefont {Solano}},
  \bibinfo {author} {\bibfnamefont {R.}~\bibnamefont {Blatt}}, \ and\ \bibinfo
  {author} {\bibfnamefont {C.~F.}\ \bibnamefont {Roos}},\ }\bibfield  {title}
  {\bibinfo {title} {\emph {{Realization of a quantum walk with one and two
  trapped ions}}},\ }\href {\doibase 10.1103/PhysRevLett.104.100503} {\bibfield
   {journal} {\bibinfo  {journal} {Physical Review Letters}\ }\textbf {\bibinfo
  {volume} {104}},\ \bibinfo {pages} {100503}\  (\bibinfo {year}
  {2010})}\BibitemShut {NoStop}%
\bibitem [{\citenamefont {Rudner}\ and\ \citenamefont
  {Levitov}(2009)}]{Rudner2009}%
  \BibitemOpen
  \bibfield  {author} {\bibinfo {author} {\bibfnamefont {M.~S.}\ \bibnamefont
  {Rudner}}\ and\ \bibinfo {author} {\bibfnamefont {L.~S.}\ \bibnamefont
  {Levitov}},\ }\bibfield  {title} {\bibinfo {title} {\emph {{Topological
  Transition in a Non-Hermitian Quantum Walk}}},\ }\href {\doibase
  10.1103/PhysRevLett.102.065703} {\bibfield  {journal} {\bibinfo  {journal}
  {Physical Review Letters}\ }\textbf {\bibinfo {volume} {102}},\ \bibinfo
  {pages} {065703}\  (\bibinfo {year} {2009})}\BibitemShut {NoStop}%
\bibitem [{\citenamefont {Kitagawa}\ \emph
  {et~al.}(2010{\natexlab{a}})\citenamefont {Kitagawa}, \citenamefont {Rudner},
  \citenamefont {Berg},\ and\ \citenamefont {Demler}}]{Kitagawa2010a}%
  \BibitemOpen
  \bibfield  {author} {\bibinfo {author} {\bibfnamefont {T.}~\bibnamefont
  {Kitagawa}}, \bibinfo {author} {\bibfnamefont {M.~S.}\ \bibnamefont
  {Rudner}}, \bibinfo {author} {\bibfnamefont {E.}~\bibnamefont {Berg}}, \ and\
  \bibinfo {author} {\bibfnamefont {E.}~\bibnamefont {Demler}},\ }\bibfield
  {title} {\bibinfo {title} {\emph {{Exploring topological phases with quantum
  walks}}},\ }\href {\doibase 10.1103/PhysRevA.82.033429} {\bibfield  {journal}
  {\bibinfo  {journal} {Physical Review A}\ }\textbf {\bibinfo {volume} {82}},\
  \bibinfo {pages} {033429}\  (\bibinfo {year}
  {2010}{\natexlab{a}})}\BibitemShut {NoStop}%
\bibitem [{\citenamefont {Rapedius}\ and\ \citenamefont
  {Korsch}(2012)}]{Rapedius2012}%
  \BibitemOpen
  \bibfield  {author} {\bibinfo {author} {\bibfnamefont {K.}~\bibnamefont
  {Rapedius}}\ and\ \bibinfo {author} {\bibfnamefont {H.~J.}\ \bibnamefont
  {Korsch}},\ }\bibfield  {title} {\bibinfo {title} {\emph
  {{Interaction-induced decoherence in non-Hermitian quantum walks of ultracold
  bosons}}},\ }\href {\doibase 10.1103/PhysRevA.86.025601} {\bibfield
  {journal} {\bibinfo  {journal} {Physical Review A}\ }\textbf {\bibinfo
  {volume} {86}},\ \bibinfo {pages} {025601}\  (\bibinfo {year}
  {2012})}\BibitemShut {NoStop}%
\bibitem [{\citenamefont {Schnyder}\ \emph {et~al.}(2009)\citenamefont
  {Schnyder}, \citenamefont {Ryu}, \citenamefont {Furusaki}, \citenamefont
  {Ludwig}, \citenamefont {Lebedev},\ and\ \citenamefont
  {Feigel’man}}]{Schnyder2008}%
  \BibitemOpen
  \bibfield  {author} {\bibinfo {author} {\bibfnamefont {A.~P.}\ \bibnamefont
  {Schnyder}}, \bibinfo {author} {\bibfnamefont {S.}~\bibnamefont {Ryu}},
  \bibinfo {author} {\bibfnamefont {A.}~\bibnamefont {Furusaki}}, \bibinfo
  {author} {\bibfnamefont {A.~W.~W.}\ \bibnamefont {Ludwig}}, \bibinfo {author}
  {\bibfnamefont {V.}~\bibnamefont {Lebedev}}, \ and\ \bibinfo {author}
  {\bibfnamefont {M.}~\bibnamefont {Feigel’man}},\ }\bibfield  {title}
  {\bibinfo {title} {\emph {{Classification of Topological Insulators and
  Superconductors}}},\ }in\ \href {\doibase 10.1063/1.3149481} {\emph {\bibinfo
  {booktitle} {AIP Conference Proceedings}}},\ Vol.\ \bibinfo {volume} {1134}\
  (\bibinfo  {publisher} {AIP},\ \bibinfo {year} {2009})\ \ pp.\ \bibinfo
  {pages} {10--21}\BibitemShut {NoStop}%
\bibitem [{\citenamefont {Asb{\'o}th}\ \emph {et~al.}(2016)\citenamefont
  {Asb{\'o}th}, \citenamefont {Oroszl{\'a}ny},\ and\ \citenamefont
  {P{\'a}lyi}}]{asboth2016short}%
  \BibitemOpen
  \bibfield  {author} {\bibinfo {author} {\bibfnamefont {J.~K.}\ \bibnamefont
  {Asb{\'o}th}}, \bibinfo {author} {\bibfnamefont {L.}~\bibnamefont
  {Oroszl{\'a}ny}}, \ and\ \bibinfo {author} {\bibfnamefont {A.}~\bibnamefont
  {P{\'a}lyi}},\ }\href {\doibase 10.1007/978-3-319-25607-8} {\emph {\bibinfo
  {title} {A Short Course on Topological Insulators [Lecture Notes in
  Physics]}}}\ (\bibinfo  {publisher} {Springer},\ \bibinfo {year}
  {2016})\BibitemShut {NoStop}%
\bibitem [{\citenamefont {Kitagawa}\ \emph
  {et~al.}(2010{\natexlab{b}})\citenamefont {Kitagawa}, \citenamefont {Berg},
  \citenamefont {Rudner},\ and\ \citenamefont {Demler}}]{Kitagawa2010}%
  \BibitemOpen
  \bibfield  {author} {\bibinfo {author} {\bibfnamefont {T.}~\bibnamefont
  {Kitagawa}}, \bibinfo {author} {\bibfnamefont {E.}~\bibnamefont {Berg}},
  \bibinfo {author} {\bibfnamefont {M.~S.}\ \bibnamefont {Rudner}}, \ and\
  \bibinfo {author} {\bibfnamefont {E.}~\bibnamefont {Demler}},\ }\bibfield
  {title} {\bibinfo {title} {\emph {{Topological characterization of
  periodically driven quantum systems}}},\ }\href {\doibase
  10.1103/PhysRevB.82.235114} {\bibfield  {journal} {\bibinfo  {journal}
  {Physical Review B}\ }\textbf {\bibinfo {volume} {82}},\ \bibinfo {pages}
  {235114}\  (\bibinfo {year} {2010}{\natexlab{b}})}\BibitemShut {NoStop}%
\bibitem [{\citenamefont {Asb{\'{o}}th}\ and\ \citenamefont
  {Obuse}(2013)}]{Asboth2013}%
  \BibitemOpen
  \bibfield  {author} {\bibinfo {author} {\bibfnamefont {J.~K.}\ \bibnamefont
  {Asb{\'{o}}th}}\ and\ \bibinfo {author} {\bibfnamefont {H.}~\bibnamefont
  {Obuse}},\ }\bibfield  {title} {\bibinfo {title} {\emph {{Bulk-boundary
  correspondence for chiral symmetric quantum walks}}},\ }\href {\doibase
  10.1103/PhysRevB.88.121406} {\bibfield  {journal} {\bibinfo  {journal}
  {Physical Review B}\ }\textbf {\bibinfo {volume} {88}},\ \bibinfo {pages}
  {121406}\  (\bibinfo {year} {2013})}\BibitemShut {NoStop}%
\bibitem [{\citenamefont {Asb{\'{o}}th}\ \emph {et~al.}(2014)\citenamefont
  {Asb{\'{o}}th}, \citenamefont {Tarasinski},\ and\ \citenamefont
  {Delplace}}]{Asboth2014}%
  \BibitemOpen
  \bibfield  {author} {\bibinfo {author} {\bibfnamefont {J.~K.}\ \bibnamefont
  {Asb{\'{o}}th}}, \bibinfo {author} {\bibfnamefont {B.}~\bibnamefont
  {Tarasinski}}, \ and\ \bibinfo {author} {\bibfnamefont {P.}~\bibnamefont
  {Delplace}},\ }\bibfield  {title} {\bibinfo {title} {\emph {{Chiral symmetry
  and bulk-boundary correspondence in periodically driven one-dimensional
  systems}}},\ }\href {\doibase 10.1103/PhysRevB.90.125143} {\bibfield
  {journal} {\bibinfo  {journal} {Physical Review B}\ }\textbf {\bibinfo
  {volume} {90}},\ \bibinfo {pages} {125143}\  (\bibinfo {year}
  {2014})}\BibitemShut {NoStop}%
\bibitem [{\citenamefont {Ruostekoski}\ \emph {et~al.}(2008)\citenamefont
  {Ruostekoski}, \citenamefont {Javanainen},\ and\ \citenamefont
  {Dunne}}]{Ruostekoski2008}%
  \BibitemOpen
  \bibfield  {author} {\bibinfo {author} {\bibfnamefont {J.}~\bibnamefont
  {Ruostekoski}}, \bibinfo {author} {\bibfnamefont {J.}~\bibnamefont
  {Javanainen}}, \ and\ \bibinfo {author} {\bibfnamefont {G.~V.}\ \bibnamefont
  {Dunne}},\ }\bibfield  {title} {\bibinfo {title} {\emph {{Manipulating atoms
  in an optical lattice: Fractional fermion number and its optical quantum
  measurement}}},\ }\href {\doibase 10.1103/PhysRevA.77.013603} {\bibfield
  {journal} {\bibinfo  {journal} {Physical Review A}\ }\textbf {\bibinfo
  {volume} {77}},\ \bibinfo {pages} {013603}\  (\bibinfo {year}
  {2008})}\BibitemShut {NoStop}%
\bibitem [{\citenamefont {Jiang}\ \emph {et~al.}(2011)\citenamefont {Jiang},
  \citenamefont {Kitagawa}, \citenamefont {Alicea}, \citenamefont {Akhmerov},
  \citenamefont {Pekker}, \citenamefont {Refael}, \citenamefont {Cirac},
  \citenamefont {Demler}, \citenamefont {Lukin},\ and\ \citenamefont
  {Zoller}}]{Jiang2011}%
  \BibitemOpen
  \bibfield  {author} {\bibinfo {author} {\bibfnamefont {L.}~\bibnamefont
  {Jiang}}, \bibinfo {author} {\bibfnamefont {T.}~\bibnamefont {Kitagawa}},
  \bibinfo {author} {\bibfnamefont {J.}~\bibnamefont {Alicea}}, \bibinfo
  {author} {\bibfnamefont {A.~R.}\ \bibnamefont {Akhmerov}}, \bibinfo {author}
  {\bibfnamefont {D.}~\bibnamefont {Pekker}}, \bibinfo {author} {\bibfnamefont
  {G.}~\bibnamefont {Refael}}, \bibinfo {author} {\bibfnamefont {J.~I.}\
  \bibnamefont {Cirac}}, \bibinfo {author} {\bibfnamefont {E.}~\bibnamefont
  {Demler}}, \bibinfo {author} {\bibfnamefont {M.~D.}\ \bibnamefont {Lukin}}, \
  and\ \bibinfo {author} {\bibfnamefont {P.}~\bibnamefont {Zoller}},\
  }\bibfield  {title} {\bibinfo {title} {\emph {{Majorana fermions in
  equilibrium and in driven cold-atom quantum wires}}},\ }\href {\doibase
  10.1103/PhysRevLett.106.220402} {\bibfield  {journal} {\bibinfo  {journal}
  {Physical Review Letters}\ }\textbf {\bibinfo {volume} {106}},\ \bibinfo
  {pages} {220402}\  (\bibinfo {year} {2011})}\BibitemShut {NoStop}%
\bibitem [{\citenamefont {Leder}\ \emph {et~al.}(2016)\citenamefont {Leder},
  \citenamefont {Grossert}, \citenamefont {Sitta}, \citenamefont {Genske},
  \citenamefont {Rosch},\ and\ \citenamefont {Weitz}}]{Leder2016a}%
  \BibitemOpen
  \bibfield  {author} {\bibinfo {author} {\bibfnamefont {M.}~\bibnamefont
  {Leder}}, \bibinfo {author} {\bibfnamefont {C.}~\bibnamefont {Grossert}},
  \bibinfo {author} {\bibfnamefont {L.}~\bibnamefont {Sitta}}, \bibinfo
  {author} {\bibfnamefont {M.}~\bibnamefont {Genske}}, \bibinfo {author}
  {\bibfnamefont {A.}~\bibnamefont {Rosch}}, \ and\ \bibinfo {author}
  {\bibfnamefont {M.}~\bibnamefont {Weitz}},\ }\bibfield  {title} {\bibinfo
  {title} {\emph {{Real-space imaging of a topological protected edge state
  with ultracold atoms in an amplitude-chirped optical lattice}}},\ }\href
  {http://arxiv.org/abs/1604.02060} {\bibfield  {journal} {\bibinfo  {journal}
  {arXiv:1604.02060}\ }\  (\bibinfo {year} {2016})}\BibitemShut {NoStop}%
\bibitem [{\citenamefont {Celi}\ \emph {et~al.}(2014)\citenamefont {Celi},
  \citenamefont {Massignan}, \citenamefont {Ruseckas}, \citenamefont {Goldman},
  \citenamefont {Spielman}, \citenamefont {Juzeliūnas},\ and\ \citenamefont
  {Lewenstein}}]{Celi2013}%
  \BibitemOpen
  \bibfield  {author} {\bibinfo {author} {\bibfnamefont {A.}~\bibnamefont
  {Celi}}, \bibinfo {author} {\bibfnamefont {P.}~\bibnamefont {Massignan}},
  \bibinfo {author} {\bibfnamefont {J.}~\bibnamefont {Ruseckas}}, \bibinfo
  {author} {\bibfnamefont {N.}~\bibnamefont {Goldman}}, \bibinfo {author}
  {\bibfnamefont {I.~B.}\ \bibnamefont {Spielman}}, \bibinfo {author}
  {\bibfnamefont {G.}~\bibnamefont {Juzeliūnas}}, \ and\ \bibinfo {author}
  {\bibfnamefont {M.}~\bibnamefont {Lewenstein}},\ }\bibfield  {title}
  {\bibinfo {title} {\emph {{Synthetic Gauge Fields in Synthetic
  Dimensions}}},\ }\href {\doibase 10.1103/PhysRevLett.112.043001} {\bibfield
  {journal} {\bibinfo  {journal} {Physical Review Letters}\ }\textbf {\bibinfo
  {volume} {112}},\ \bibinfo {pages} {043001}\  (\bibinfo {year}
  {2014})}\BibitemShut {NoStop}%
\bibitem [{\citenamefont {Mancini}\ \emph {et~al.}(2015)\citenamefont
  {Mancini}, \citenamefont {Pagano}, \citenamefont {Cappellini}, \citenamefont
  {Livi}, \citenamefont {Rider}, \citenamefont {Catani}, \citenamefont {Sias},
  \citenamefont {Zoller}, \citenamefont {Inguscio}, \citenamefont {Dalmonte},\
  and\ \citenamefont {Fallani}}]{Mancini}%
  \BibitemOpen
  \bibfield  {author} {\bibinfo {author} {\bibfnamefont {M.}~\bibnamefont
  {Mancini}}, \bibinfo {author} {\bibfnamefont {G.}~\bibnamefont {Pagano}},
  \bibinfo {author} {\bibfnamefont {G.}~\bibnamefont {Cappellini}}, \bibinfo
  {author} {\bibfnamefont {L.}~\bibnamefont {Livi}}, \bibinfo {author}
  {\bibfnamefont {M.}~\bibnamefont {Rider}}, \bibinfo {author} {\bibfnamefont
  {J.}~\bibnamefont {Catani}}, \bibinfo {author} {\bibfnamefont
  {C.}~\bibnamefont {Sias}}, \bibinfo {author} {\bibfnamefont {P.}~\bibnamefont
  {Zoller}}, \bibinfo {author} {\bibfnamefont {M.}~\bibnamefont {Inguscio}},
  \bibinfo {author} {\bibfnamefont {M.}~\bibnamefont {Dalmonte}}, \ and\
  \bibinfo {author} {\bibfnamefont {L.}~\bibnamefont {Fallani}},\ }\bibfield
  {title} {\bibinfo {title} {\emph {{Observation of chiral edge states with
  neutral fermions in synthetic Hall ribbons}}},\ }\href {\doibase
  10.1126/science.aaa8736} {\bibfield  {journal} {\bibinfo  {journal}
  {Science}\ }\textbf {\bibinfo {volume} {349}},\ \bibinfo {pages} {1510}\
  (\bibinfo {year} {2015})}\BibitemShut {NoStop}%
\bibitem [{\citenamefont {Stuhl}\ \emph {et~al.}(2015)\citenamefont {Stuhl},
  \citenamefont {Lu}, \citenamefont {Aycock}, \citenamefont {Genkina},\ and\
  \citenamefont {Spielman}}]{Stuhl2015}%
  \BibitemOpen
  \bibfield  {author} {\bibinfo {author} {\bibfnamefont {B.~K.}\ \bibnamefont
  {Stuhl}}, \bibinfo {author} {\bibfnamefont {H.~I.}\ \bibnamefont {Lu}},
  \bibinfo {author} {\bibfnamefont {L.~M.}\ \bibnamefont {Aycock}}, \bibinfo
  {author} {\bibfnamefont {D.}~\bibnamefont {Genkina}}, \ and\ \bibinfo
  {author} {\bibfnamefont {I.~B.}\ \bibnamefont {Spielman}},\ }\bibfield
  {title} {\bibinfo {title} {\emph {{Visualizing edge states with an atomic
  Bose gas in the quantum Hall regime}}},\ }\href {\doibase
  10.1126/science.aaa8515} {\bibfield  {journal} {\bibinfo  {journal}
  {Science}\ }\textbf {\bibinfo {volume} {349}},\ \bibinfo {pages} {1514}\
  (\bibinfo {year} {2015})}\BibitemShut {NoStop}%
\bibitem [{\citenamefont {Creutz}(1999)}]{Creutz1999}%
  \BibitemOpen
  \bibfield  {author} {\bibinfo {author} {\bibfnamefont {M.}~\bibnamefont
  {Creutz}},\ }\bibfield  {title} {\bibinfo {title} {\emph {{End States, Ladder
  Compounds, and Domain-Wall Fermions}}},\ }\href {\doibase
  10.1103/PhysRevLett.83.2636} {\bibfield  {journal} {\bibinfo  {journal}
  {Physical Review Letters}\ }\textbf {\bibinfo {volume} {83}},\ \bibinfo
  {pages} {2636}\  (\bibinfo {year} {1999})}\BibitemShut {NoStop}%
\bibitem [{\citenamefont {Bermudez}\ \emph {et~al.}(2009)\citenamefont
  {Bermudez}, \citenamefont {Patan{\`{e}}}, \citenamefont {Amico},\ and\
  \citenamefont {Martin-Delgado}}]{Bermudez2009}%
  \BibitemOpen
  \bibfield  {author} {\bibinfo {author} {\bibfnamefont {A.}~\bibnamefont
  {Bermudez}}, \bibinfo {author} {\bibfnamefont {D.}~\bibnamefont
  {Patan{\`{e}}}}, \bibinfo {author} {\bibfnamefont {L.}~\bibnamefont {Amico}},
  \ and\ \bibinfo {author} {\bibfnamefont {M.~A.}\ \bibnamefont
  {Martin-Delgado}},\ }\bibfield  {title} {\bibinfo {title} {\emph
  {{Topology-Induced Anomalous Defect Production by Crossing a Quantum Critical
  Point}}},\ }\href {\doibase 10.1103/PhysRevLett.102.135702} {\bibfield
  {journal} {\bibinfo  {journal} {Physical Review Letters}\ }\textbf {\bibinfo
  {volume} {102}},\ \bibinfo {pages} {135702}\  (\bibinfo {year}
  {2009})}\BibitemShut {NoStop}%
\bibitem [{\citenamefont {Sticlet}\ \emph {et~al.}(2014)\citenamefont
  {Sticlet}, \citenamefont {Seabra}, \citenamefont {Pollmann},\ and\
  \citenamefont {Cayssol}}]{Sticlet2014}%
  \BibitemOpen
  \bibfield  {author} {\bibinfo {author} {\bibfnamefont {D.}~\bibnamefont
  {Sticlet}}, \bibinfo {author} {\bibfnamefont {L.}~\bibnamefont {Seabra}},
  \bibinfo {author} {\bibfnamefont {F.}~\bibnamefont {Pollmann}}, \ and\
  \bibinfo {author} {\bibfnamefont {J.}~\bibnamefont {Cayssol}},\ }\bibfield
  {title} {\bibinfo {title} {\emph {{From fractionally charged solitons to
  Majorana bound states in a one-dimensional interacting model}}},\ }\href
  {\doibase 10.1103/PhysRevB.89.115430} {\bibfield  {journal} {\bibinfo
  {journal} {Physical Review B}\ }\textbf {\bibinfo {volume} {89}},\ \bibinfo
  {pages} {115430}\  (\bibinfo {year} {2014})}\BibitemShut {NoStop}%
\bibitem [{\citenamefont {Jackiw}\ and\ \citenamefont
  {Rebbi}(1976)}]{Jackiw1976}%
  \BibitemOpen
  \bibfield  {author} {\bibinfo {author} {\bibfnamefont {R.}~\bibnamefont
  {Jackiw}}\ and\ \bibinfo {author} {\bibfnamefont {C.}~\bibnamefont {Rebbi}},\
  }\bibfield  {title} {\bibinfo {title} {\emph {{Solitons with fermion
  number}}},\ }\href {\doibase 10.1103/PhysRevD.13.3398} {\bibfield  {journal}
  {\bibinfo  {journal} {Physical Review D}\ }\textbf {\bibinfo {volume} {13}},\
  \bibinfo {pages} {3398}\  (\bibinfo {year} {1976})}\BibitemShut {NoStop}%
\bibitem [{\citenamefont {Haake}(2010)}]{Haake2010}%
  \BibitemOpen
  \bibfield  {author} {\bibinfo {author} {\bibfnamefont {F.}~\bibnamefont
  {Haake}},\ }\href {\doibase 10.1007/978-3-642-05428-0} {\emph {\bibinfo
  {title} {Quantum Signatures of Chaos}}}\ (\bibinfo  {publisher}
  {Springer-Verlag Berlin Heidelberg},\ \bibinfo {address} {Berlin,
  Heidelberg},\ \bibinfo {year} {2010})\BibitemShut {NoStop}%
\end{thebibliography}%

\begin{appendix}

\section{mapping to the Creutz ladder}
\label{sec:MapToCreutz}

In this section, we will show that, under a simple change of basis, the system maps exactly onto a well known topologically non-trivial 1D system, the Creutz ladder \cite{Creutz1999, Bermudez2009, Sticlet2014}. We will then use this knowledge to re-derive the phase diagram shown in Fig.\ \ref{fig:phase_diagram}.

The Creutz ladder describes a spinless particle hopping in a 1D ladder, as sketched in Fig.\ \ref{fig:Creutz_ladder}. The particularity of this model is that the particle has amplitude to hop in the diagonal directions. We have chosen the basis of spin space such that Fig.\ \ref{fig:Creutz_ladder} is reminiscent of Fig.\ \ref{fig:tunnelling_sketches}(b).

\begin{figure}[t]
\centering
\includegraphics[width=0.35\textwidth]{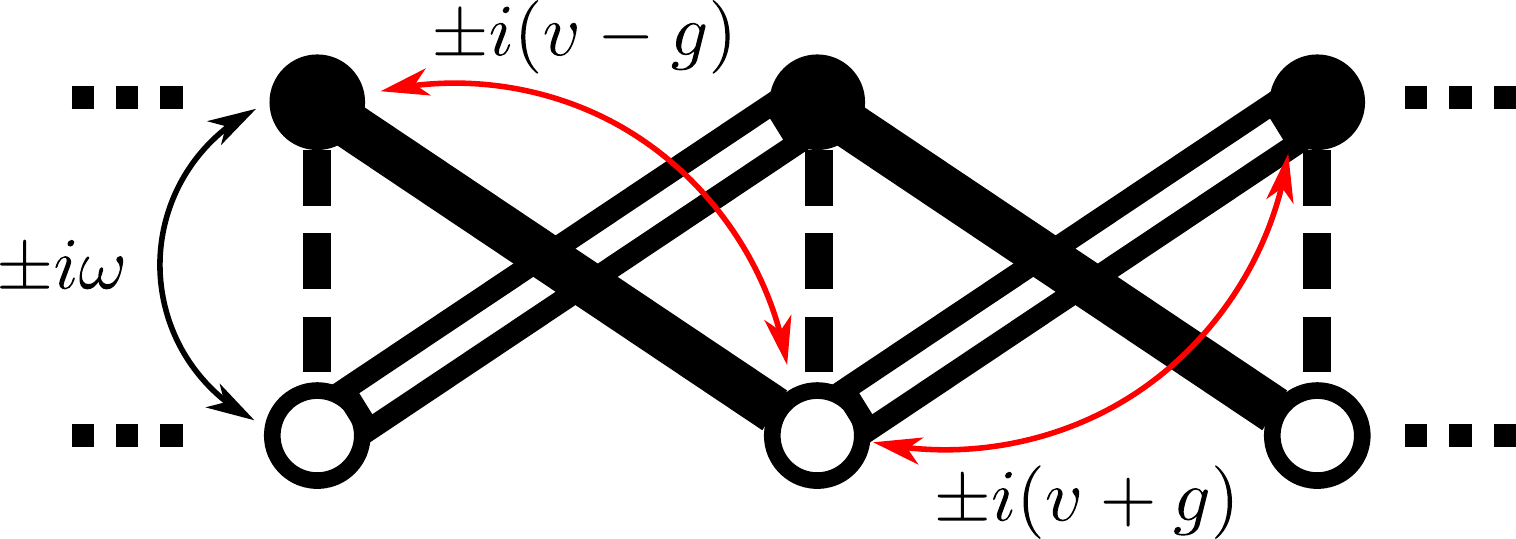}
\caption{1D Creutz ladder. The particle has amplitude $\pm i w$ of tunnelling along the rungs of the ladder (dashed line), and diagonal tunnelling amplitudes $\pm i(v-g)$ (solid line) and $\pm i(v+g)$ (double line). The tunnelling amplitudes are indicated by curved arrows, the colour of which is unimportant. Sites belonging to the upper (lower) part of the ladder are indicated by full (empty) circles. This system is topologically non-trivial, with winding number given by Eq.\ (\ref{eq:Creutz_winding}).}
\label{fig:Creutz_ladder}
\end{figure}

The Hamiltonian of the Creutz ladder has the form:
\begin{equation}
\label{eq:H_Creutz}
\text{\^H}_C = \frac{1}{2} \sum_n \cc_{n}^\dagger w  \sigma_2 \cc_n +\cc_{n+1}^\dagger(i v  \sigma_1-g \sigma_2)\cc_n + \text{H.c}).
\end{equation}
As previously, $\cc^\dagger_n$ ($\cc_n$) creates (annihilates) a particle with two internal states on site $n$. The $\sigma_i$ matrices denotes the Pauli matrices acting in the space of sites perpendicular to the axis of the ladder (represented in the vertical direction in Fig.\ \ref{fig:Creutz_ladder}), with $i\in\{1,  2,  3\}$. We will assume that the hopping amplitudes $v$, $w$ and $g$ are all real parameters, and  block-diagonalise Eq.\ (\ref{eq:H_Creutz}) by Fourier transformation:
\begin{equation}
\label{eq:H_Creutz_Kspace}
\text{\^H}_C(k) = -v \sin(k d)\sigma_1+(w-g \cos(k d))  \sigma_2,
\end{equation}
where we have set the ladder's unit cell size to $d$. The Creutz ladder belongs to the BDI class of the topological classification of Hamiltonians \cite{Sticlet2014}. As a result, this system admits a non-zero winding number  $\nu_C$ , the value of which depends on the system's parameters as:
\begin{equation}
\label{eq:Creutz_winding}
\nu_C=\frac{1}{2}(\text{sign}(w+g)-\text{sign}(w-g)).
\end{equation}

We will now show that we can map the atomic quantum walk to the Creutz ladder. We consider the translational invariant atomic quantum walk in the two-state basis presented in Sec.\ \ref{sec:Model}, and change the origin of momentum $k\rightarrow k+\pi/(2 d)$. In this basis, Eq.\ (\ref{eq:Ht'}) becomes:
\begin{equation}
\label{eq:Ht''}
\text{\^H}_S'(k)\rightarrow \text{\~H}_S(k) = -2  J\sin(k d) \sigma_1 + 2 \delta\cos(k d) \sigma_2.
\end{equation}
The Hamiltonian \^H$_\theta'$ given by Eq.\ (\ref{eq:Htheta'}) is not modified by this transformation. Note that this is a trivial Gauge transformation that cannot change the topological properties of the system. Interestingly, in this basis, the system has CS, TRS and PHS. The operators implementing these symmetries are detailed in the Table \ref{tab:AQW_Sym}. The method used to determine the system's symmetries is detailed in the Appendix \ref{sec:Symmetries}.

\begin{table}[t]
\centering
\caption{In the first column, the operators implementing various symmetries of \^H$_F'$ are listed. We list the squares of these operators in the second column.}
\label{tab:AQW_Sym}
\begin{tabular}{ |c|c|c| }
\hline
symmetry & operator & square\\
\hline
Chiral symmetry & $\sigma_3$ & $\sigma_0$ \\
Time reversal symmetry & $\sigma_3  \hat{\mathcal{K}}$ & $\sigma_0$ \\
Particle hole symmetry & $\hat{\mathcal{K}}$ & $\sigma_0$ \\
\hline
\end{tabular}
\end{table}

From Table \ref{tab:AQW_Sym}, we find that the system presents CS, TRS and PHS, all of which square to the identity. This tells us immediately that we are in the BDI class of the topological classification of Hamiltonians, i.e the same symmetry class as the Creutz ladder.

We will now proceed to show that \~H$_F$, the Floquet Hamiltonian, maps onto the Creutz ladder when $J=\delta$. We can find \~H$_F$ by substituting Eqs.\ (\ref{eq:Htheta'}) and (\ref{eq:Ht''}) into Eq.\ (\ref{eq:U'}). In the limit $J=\delta$, \~{H}$_F$ is:
\begin{equation}
\begin{split}
& \text{\~H}_F= 
\pm\frac{\tilde{E}(k)}{\sin(\tilde{E}(k))}\Big(-\text{sign}(\delta)\sin(2\delta)\sin(k d) \sigma_1\\
& +(-\cos(2\delta)\sin(\theta)+\text{sign}(\delta) \cos(\theta)\sin(2\delta)\cos(k d)) \sigma_2\Big).
\end{split}
\end{equation}
Apart from the upfront $\tilde{E}(k)/ \sin(\tilde{E}(k))$, this is exactly the Creutz ladder Hamiltonian Eq.\ (\ref{eq:H_Creutz_Kspace}), with:
\begin{eqnarray}
\label{eq:CreutzTunnelling1}
v & = & \text{sign}(\delta)\sin(2\delta),\\
\label{eq:CreutzTunnelling2}
w & = & -\text{sign}(\delta)\cos(2\delta)\sin(\theta),\\
\label{eq:CreutzTunnelling3}
g & = & -\cos(\theta)\sin(2\delta).
\end{eqnarray}
By substituting these values in Eq.\ (\ref{eq:Creutz_winding}), we can calculate the winding number $\nu_C$ and deduce the phase diagram in this time frame, which is represented in Fig.\ \ref{fig:Creutz_phase_diagram}. While we expect the upfront term $\tilde{E}(k)/ \sin(\tilde{E}(k))$ to deform the band structure, it does not change the symmetry properties of \~H$_F$, and therefore cannot change its topological properties as long as it does not close the band gap.

\begin{figure}[t]
\centering
\includegraphics[width=0.42\textwidth]{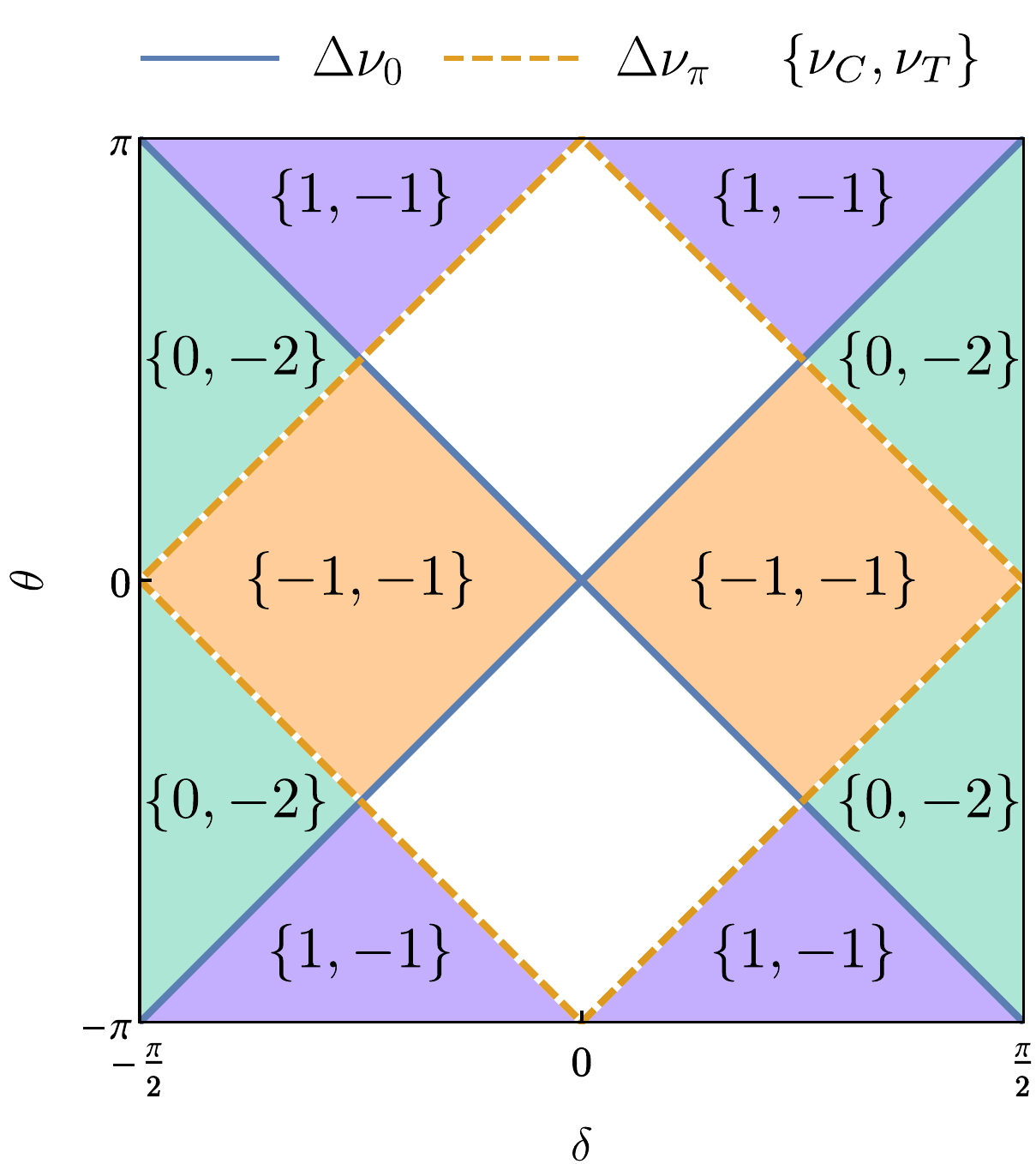}
\caption{Winding numbers $\{\nu_C,\nu_T\}$ of the atomic quantum walk when $J=\delta$. In the time-frame Eq.\ \eqref{eq:U'}, the system maps onto the Creutz ladder and $\nu_C$ is accurately predicted by Eq.\ (\ref{eq:Creutz_winding}). $\nu_T$ is the winding number in the other symmetric time frame Eq.\ \eqref{eq:timeframe2}. These are related to the winding numbers $\nu_0,\nu_\pi$ through Eqs.\ \eqref{eq:nuC} and \eqref{eq:nuT}.}
\label{fig:Creutz_phase_diagram}
\end{figure}

In expressing \^U$'$ in the symmetric form for Eq.\ \eqref{eq:U'}, we made a choice of time frame. There exists another time frame which has an inversion point in time and has the form:
\begin{equation}
\label{eq:timeframe2}
\text{\^U}_T'=e^{-i\Hs'/2}e^{-i\Htheta'}e^{-i\Hs'/2}.
\end{equation}
While the system in this time frame does not map onto the Creutz ladder, it does have a winding number $\nu_{T}$, which is the second topological invariant represented on Fig.\ \ref{fig:Creutz_phase_diagram}. By comparing Figs.\ \ref{fig:phase_diagram} and \ref{fig:Creutz_phase_diagram} when $\delta>0$, we notice that:
\begin{align}
\label{eq:nuC}
& \nu_C=\nu_0+\nu_\pi,\\
\label{eq:nuT}
& \nu_{T}=\nu_0-\nu_\pi,
\end{align}
in agreement with Ref.\ \cite{Asboth2014}. When $J<0$ all winding numbers from Fig.\ \ref{fig:phase_diagram} change sign. Taking this into account, we see that the winding numbers from Fig.\ \ref{fig:Creutz_phase_diagram} obey Eqs.\ \eqref{eq:nuC} and \eqref{eq:nuT} also when $J=\delta<0$. Remember that the phase of the atomic quantum walk is independent of the absolute value of $J$. Thus the Creutz ladder accurately gives the winding number of the atomic quantum walk even when $J\neq \delta$.

\section{double atomic quantum walk}
\label{sec:DoubleQW}

As we explained in section \ref{sec:Model}, the system has the dynamics of a quantum walk. As we mentioned in this same section however, our protocol is slightly more complex than the standard quantum walk considered in Refs.\ \cite{Karski2009, Kitagawa2010a, Robens2015a}. Indeed, we saw that each eigenstate of $\Hs$ is doubly degenerate, implying that two wavepackets exist in each circled region of Fig.\ \ref{fig:Ht_dispersion}. As a result of this, our quantum walk has four distinct walkers, and the coin operation $\CCC$ can be shown to couple all four. In the following, we will start by mapping the system in the simplified (two state) basis to a standard quantum walk. We then show that the system which we simulated in Sec.\ \ref{sec:Model} has in reality the dynamics of two independent quantum walks.

\begin{figure}[t]
\centering
\includegraphics[width=0.95\columnwidth]{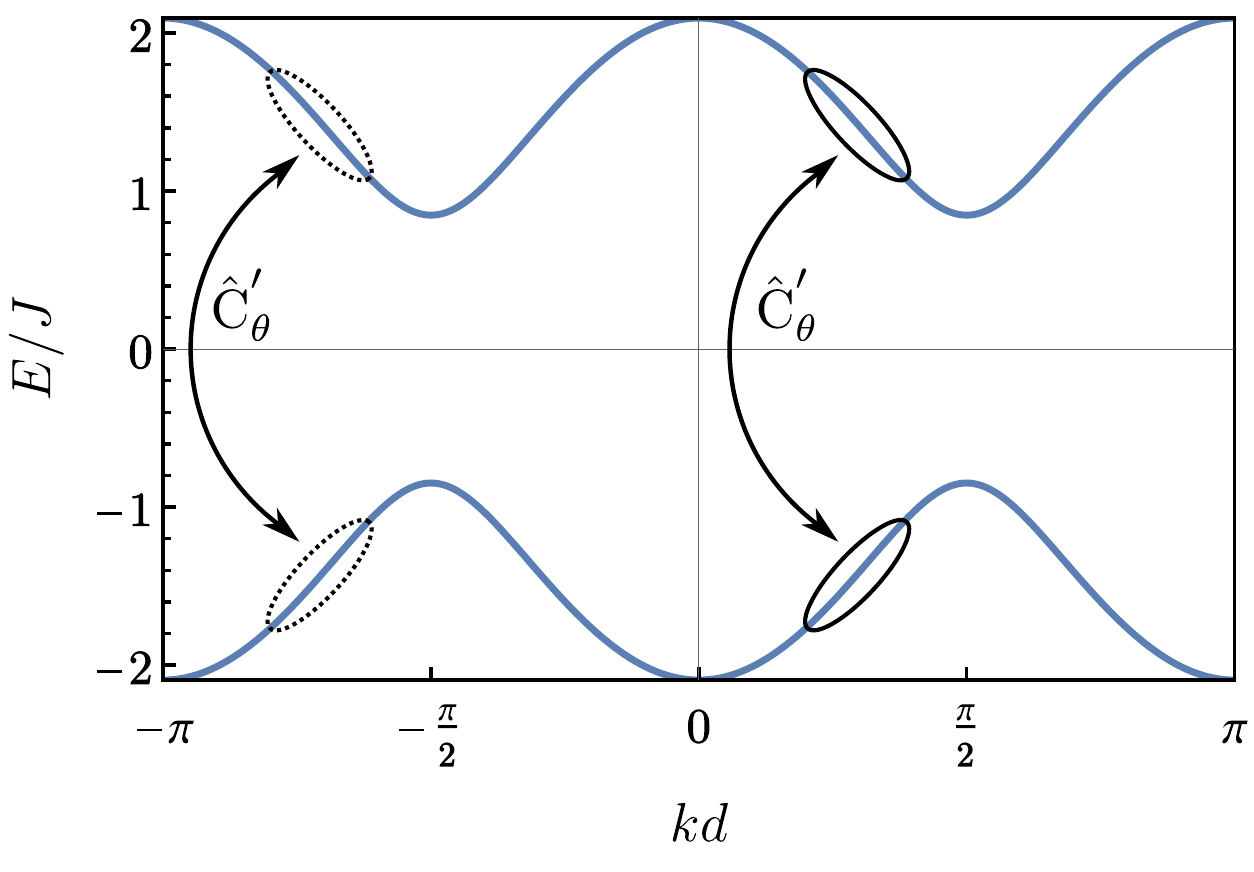}
\caption{Dispersion of \^H$_S'$ Eq.\ (\ref{eq:Ht_Kspace}) for $J=\pi/3$ and $\delta=0.42$. The wavepackets centred around $k=\pi/(4d)$, where the slope is locally linear, are circled by a solid line. Due to the symmetry of the dispersion about $E=0$, these states move at equal and opposite average velocities. The regions circled in Fig.\ \ref{fig:Ht_dispersion} correspond to a superposition of the wavepackets centred at $k=\pi/(4d)$ and the ones centred at $k=-3\pi/(4d)$ (circled by a dotted line). The couplings induced by \^C$_\theta'$ are illustrated by arrows. When both the states in the region of $k=\pi/(4d)$ and $k=-3\pi/(4d)$ are populated, the system performs two simultaneous, independent quantum walks.}
\label{fig:Hs'_dispersion}
\end{figure}

The Hamiltonian $\Hs'$ Eq.\ \eqref{eq:Ht_Kspace} can readily be diagonalised, yielding the eigenstates:
\begin{align}
& \ket{\pm',k}=2^{-1/2}(1,\pm e^{i\phi(k)}),\\
& \phi(k)=\arg(J \cos(kd)+i\delta\sin(kd)),
\end{align}
where $\ket{+',k},\ket{-',k}$ are the eigenstates belonging to the top and bottom bands respectively, and ``arg'' denotes the phase of the complex number. From this definition, we define a pair of walkers in this basis as wavepackets centred narrowly around $k=\pi/(4d)$, such that the left- and right-walker belong to the top and bottom bands respectively, as represented on Fig.\ \ref{fig:Hs'_dispersion}. As previously, thanks to the symmetry of the spectrum about $E=0$, these are translated in real space by $\Hs'$ at equal and opposite average velocities.

To understand in what sense this system describes a quantum
walk, we express $\CCC'$ in the basis $\ket{+',k}, \ket{-',k}$:
\begin{align}
\CCC' &= e^{-i\Htheta'};\\
\text{H}_\theta' &= \theta
\begin{pmatrix}
\sin(\phi(k)) & i \cos(\phi(k)) \\
-i \cos(\phi(k)) & -\sin(\phi(k))
\end{pmatrix}.
\end{align}
This corresponds to a rotation by an angle $\theta$ about the axis $\left(0,-\cos(\phi(k)),\sin(\phi(k))\right)$.

To recapitulate, we have defined left- and right-walkers, which are translated in real space at average equal and opposite velocities. Our walkers are periodically coupled by the $\CCC'$ operation, as indicated in Fig.\ \ref{fig:Hs'_dispersion}. In this sense the protocol fits exactly our definition of a quantum walk. Note however that no choice of $\theta$ can in general fully interchange right- and left-walkers; in this respect it is different from the quantum walks considered in Refs.\ \cite{Karski2009, Kitagawa2010a, Robens2015a}.

An additional subtlety comes from the choice of initial state. Let $\ket{+,k,\uparrow},\ket{+,k,\downarrow}$ be the eigenstates of $\Hs$ Eq.\ \eqref{eq:Ht} in the superlattice basis, corresponding to the state with quasimomentum $k$ belonging to the top band with spin $\uparrow$ or $\downarrow$ respectively, and $\ket{-,k,\uparrow},\ket{-,k,\downarrow}$ their bottom band counterparts. In general, these states are related to the eigenstates of $\Hs'$ through:
\begin{align}
2\ket{\pm,k,\uparrow} & = \left(\tau_0+\tau_3\right)\otimes \left(\ket{\pm',k}+ \ket{\pm',k+\pi}\right)\\
& \left(\tau_0-\tau_3\right)\otimes\sigma_1 \cdot \left(\ket{\pm',k}- \ket{\pm',k+\pi}\right),
\end{align}
and
\begin{align}
2\ket{\pm,k,\downarrow} & = \left(\tau_0+\tau_3\right)\otimes \left(\ket{\pm',k}- \ket{\pm',k+\pi}\right)\\
& \left(\tau_0-\tau_3\right)\otimes\sigma_1 \cdot \left(\ket{\pm',k}+ \ket{\pm',k+\pi}\right),
\end{align}
where only the $\sigma_i$ matrices act on the states $\ket{\pm',k}$.

When simulating the atomic quantum walk, we chose as an initial state the Gaussian wavepacket narrowly centred around $k=\pi/(4d)$ in the superlattice basis, as circled on Fig.\ \ref{fig:Ht_dispersion}. This state can be expressed in the simpler basis of $\Hs'$ and $\CCC'$ as a superposition of wavepackets centred about $k=\pi/(4d)$ and $k=-3\pi/(4d)$, as represented on Fig.\ \ref{fig:Hs'_dispersion} by solid and dotted circles respectively. With this choice of initial state, we are performing a quantum walk with four walkers (two right- and two left-walkers, see Fig.\ \ref{fig:Hs'_dispersion}). As states which have different quasimomentum are not coupled by $\CCC'$, it is clear that two independent quantum walks are simultaneously being performed. This, however, should have no effect on the outcome of the simulation, as the slope at $k=\pi/(4d)$ and $k=-3\pi/(4d)$ is exactly the same for any choice of parameters.

\section{symmetries of the atomic quantum walk}
\label{sec:Symmetries}

In general in 1D, a system can display non-trivial topological behaviour only if it is constrained by certain symmetries. We are interested in whether or not the atomic quantum walk presents time reversal symmetry (TRS), particle hole symmetry (PHS) and chiral symmetry (CS), which determine the system's topological class. While we study the case specific to our system, a more general and complete study of symmetries and their relevance to topological phases can be found in Ref.\ \cite{Haake2010}.
For simplicity, we will work in the basis where \^U$'$ is given by Eq.\ \eqref{eq:U'}

The system has CS if there is a unitary operator $\hat{\Gamma}$ acting within a single unit cell which anti-commutes with the Hamiltonian:
\begin{equation}
\label{eq:AQW_ChiralSymm}
\hat{\Gamma} \cdot \hat{\text{H}}_F'(k) \cdot \hat{\Gamma}^\dagger = -\hat{\text{H}}_F'(k) ~ \Rightarrow ~ \hat{\Gamma} \cdot \hat{\text{U}}'(k) \cdot \hat{\Gamma}^\dagger = \hat{\text{U}}'(k)^\dagger.
\end{equation}
Thanks to the symmetric form of Eq.\ \eqref{eq:U'}, if an operator $\hat{\Gamma}$ simultaneously anti-commutes with \^H$_\theta'$ and \^H$_S'$, it automatically satisfies Eq.\ (\ref{eq:AQW_ChiralSymm}):
\begin{equation}
\begin{split}
\text{\^U}'^\dagger & = e^{i \text{\^H}_\theta'/2}\cdot e^{i \text{\^H}_S'} \cdot e^{i \text{\^H}_\theta'/2}\\
\Rightarrow \hat{\Gamma}\cdot\text{\^U}'^\dagger\cdot\hat{\Gamma}^\dagger & = e^{i \hat{\Gamma}\cdot\text{\^H}_\theta'\cdot\hat{\Gamma}^\dagger /2}\cdot e^{i \hat{\Gamma}\cdot\text{\^H}_S'\cdot\hat{\Gamma}^\dagger} \cdot e^{i \hat{\Gamma}\cdot\text{\^H}_\theta'\cdot\hat{\Gamma}^\dagger /2}\\
& = \text{\^U}'.
\end{split}
\end{equation}
Thus, if $\hat{\Gamma}$ is a valid CS for \^H$_\theta'$ and \^H$_S'$, it is also a CS operator for \^U$'$. By inspection of Eqs.\ (\ref{eq:Htheta'}) and (\ref{eq:Ht_Kspace}), we find that $\hat{\Gamma}=\sigma_3$ simultaneously anti-commutes with \^H$_\theta'$ and \^H$_S'$, and is therefore the CS operator in this basis.

But is it the only operator which satisfies Eq.\ (\ref{eq:AQW_ChiralSymm})? Given a matrix \^U$'$ which has the form Eq.\ (\ref{eq:U'}), it can happen that $\hat{\Gamma}$ satisfies Eq.\ (\ref{eq:AQW_ChiralSymm}) without simultaneously anti-commuting with \^H$_S'$ and \^H$_\theta'$. Assuming this is true, however, leads to strong constraints on the form of \^H$_S'$ and \^H$_\theta'$. In particular, when \^H$_S'$ is a function of quasimomentum (as in the present case), there will in general exist no additional CS operator. This result must remain valid when $\theta$ varies spatially. Indeed,  breaking translational invariance cannot introduce new symmetries in the system.

We can now turn to the other symmetries of the system, starting with TRS. The system has TRS symmetry if there is an anti-unitary operator $\hat{\mathcal{T}}$ which commutes with the Hamiltonian, and acts only within a single unit cell. Without loss of generality, we can express $\hat{\mathcal{T}}$ as the product of a unitary operator $\hat{\tau}$ and $\hat{\mathcal{K}}$, the complex conjugation operator: $\hat{\mathcal{T}}=\hat{\tau}  \hat{\mathcal{K}}$. The complex conjugation operator is an anti-unitary operator which acts as:
\begin{equation}
\langle \hat{\mathcal{K}} n | \hat{\mathcal{K}} \psi\rangle = \psi(n)^*.
\end{equation}
Using $|k\rangle=\sum_n \exp(-iknd)|n\rangle$, we find that:
\begin{equation}
\langle \hat{\mathcal{K}} k | \hat{\mathcal{K}} \psi\rangle =
\sum_n e^{-iknd} \psi(n)^*
=\psi(-k)^*.
\end{equation}
Thus $\hat{\mathcal{K}}$ sends $k\rightarrow -k$. Searching for a TRS operator therefore amounts to finding $\hat{\tau}$ such that:
\begin{equation}
\label{eq:AQW_TRS}
\hat{\tau} \cdot \hat{\text{H}}_F'(-k)^T \cdot \hat{\tau}^\dagger = \hat{\text{H}}_F'(k) ~ \Rightarrow ~ \hat{\tau} \cdot \hat{\text{U}}'(-k)^T \cdot \hat{\tau}^\dagger = \hat{\text{U}}'(k).
\end{equation}
As previously, if an anti-unitary $\hat{\mathcal{T}}$ simultaneously commutes with \^H$_S'(k)$ and \^H$_\theta'$, it automatically satisfies Eq.\ (\ref{eq:AQW_TRS}). In this case however, no such operator exists. This is not sufficient to say that the system does not have TRS. As was the case with CS however, the existence of $\hat{\mathcal{T}}$ which satisfies Eq.\ (\ref{eq:AQW_TRS}) without simultaneously commuting with \^H$_S'(k)$ and \^H$_\theta'$ would imply strong constraints on these matrices. These are in general not satisfied when \^H$_S'(k)$ and \^H$_\theta'$ are functions of independent variables. We can confirm numerically that the system does not have TRS by plotting the dispersion of \^H$_F'$. Indeed, TRS implies that any eigenstate of \^H$_F'$ has a partner eigenstate with equal energy and opposite quasimomentum. Because the system's spectrum is not symmetric about $k=0$, we can conclude that the system does not present TRS.

Finally, the system has PHS symmetry if there is a anti-unitary operator $\hat{\mathcal{P}}$ which anti-commutes with the Hamiltonian. We define $\hat{\varrho}$, the unitary part of $\hat{\mathcal{P}}$, such that: $\hat{\mathcal{P}}=\hat{\varrho}  \hat{\mathcal{K}}$. This operator satisfies:
\begin{equation}
\label{eq:AQW_PHS}
\hat{\varrho} \cdot \hat{\text{H}}_F(-k)^T \cdot \hat{\varrho}^\dagger = -\hat{\text{H}}_F(k) ~ \Rightarrow ~ \hat{\varrho} \cdot \hat{\text{U}}(-k)^T \cdot \hat{\varrho}^\dagger = \hat{\text{U}}(k)^\dagger.
\end{equation}
We already know that there is no such operator due to the absence of TRS. Indeed, if \^U$'$ admitted both CS and PHS, their product would yield an anti-unitary matrix which commutes with the Hamiltonian, and this operator would satisfy Eq.\ \eqref{eq:AQW_TRS}. As no such operator exists, we can conclude that PHS is also absent from this system.

The presence of only chiral symmetry implies that the atomic quantum walk belongs to the AIII class of the classification of topological phases. Hamiltonians in this symmetry class can in general have non-zero winding numbers \cite{Schnyder2008}.

As a closing remark, we remind the reader that in Sec.\ \ref{sec:Model}, we approximated the time evolution by Eq.\ (\ref{eq:U}). We were able to do this by saying that the spin mixing pulse is so short and intense that \^H$_S$ is negligible during this period. We now point out that, because the CS operator anti-commutes simultaneously with \^H$_S'$ and \^H$_\theta'$, even when our approximation breaks down, CS is not broken. Finally, note that $\hat{\Gamma}=\sigma_3$ is an operator which acts within a single unit cell. This implies that we can break translational invariance without breaking chiral symmetry.

\end{appendix}

\end{document}